\documentclass[aps,pre,amssymb,reprint]{revtex4-1}
\usepackage{amsmath,amsfonts,amsthm,bm}
\usepackage{graphicx}
\usepackage{subcaption}

\begin{document}
\title{Implicit Molecular Stresses in Weakly-Compressible Particle-Based Discretization Methods for Fluid Flow}
\author{Max Okraschevski}
\email{E-Mail: max.okraschevski@kit.edu}
\author{Niklas Buerkle}
\author{Rainer Koch}
\author{Hans-Joerg Bauer}
\affiliation{Institute of Thermal Turbomachinery, Karlsruhe Institute of Technology, Kaiserstraße 12, 76131 Karlsruhe, Germany}

\begin{abstract}
Weakly-compressible particle-based discretization methods, utilized for the solution of the subsonic Navier-Stokes equation, are gaining increasing popularity in the fluid dynamics community. One of the most popular among these methods is the weakly-compressible smoothed particle hydrodynamics (WCSPH). Since the dynamics of a single numerical particle is determined by fluid dynamic transport equations, the particle per definition should represent a homogeneous fluid element. However, it can be easily argued that a single particle behaves only pseudo-Lagrangian as it is affected by volume partition errors and can hardly adapt its shape to the actual fluid flow. Therefore, we will assume that the kernel support provides a better representative of an actual fluid element. By means of non-equilibrium molecular dynamics (NEMD) analysis, we derive isothermal transport equations for a kernel-based fluid element. The main discovery of the NEMD analysis is a molecular stress tensor which may serve to explain current problems encountered in applications of weakly-compressible particle-based discretization methods.
\end{abstract}

\maketitle

\section{Introduction}
\label{sec:Motivation}
Nowadays, the weakly-compressible smoothed particle hydrodynamics (WCSPH) method is a popular method utilized for solving the subsonic Navier-Stokes equation. Within the method, the fluid domain is decomposed into a finite set of constant mass particles. These particles interact with each other by means of a compact, spherical kernel. The dynamics of each particle is determined by transport equations in a Lagrangian frame of reference. Hence, the particles should be representatives of real fluid elements per definition. Unfortunately, their performance to represent the latter is insufficient. As explained by Vogelsberger et al. \cite{Vogelsberger_2012}, this is due to a conceptual problem of SPH which causes the method to behave only pseudo-Lagrangian at finite resolution. Since we believe that the pseudo-Lagrangian character is the physical root of well-known problems in the community, especially excessive numerical dissipation \cite{Colagrossi_2013}, we strive for a mathematical description of it. To date, such a measure does not exist to the author's knowledge. Within the paper, this problem will be tackled. Since the pseudo-Lagrangian character is fundamental for our theory, we start with a description of its origin which can be attributed to two reasons. \\
On the one hand, the pseudo-Lagrangian character is introduced by the common low-order particle volume estimate $V$ as initially suggested by Espa\~{n}ol and Revenga \cite{Espanol_2003}:
\begin{equation}
\frac{1}{V} \approx \sum_{j=1}^{N_{NGB}} W_h(\mathbf{x}-\mathbf{x}_j)~.
\label{eq:VolEst}
\end{equation}
In Eq. (\ref{eq:VolEst}), the function $W_h(\mathbf{x}-\mathbf{x}_j)$ is a spherical, compact kernel with $\mathbf{x} \in \mathbb{R}^3$ denoting the center particle position and $\mathbf{x}_j \in \mathbb{R}^3$ the neighbor particle positions. The parameter $h$ represents the smoothing length which defines the local, limiting radius of interaction between particles. $N_{NGB}$ is the number of particles located within the kernel support. In this work, $h$ is set constant for the sake of simplicity. 
The problem with the approximation of $V$ by Eq. (\ref{eq:VolEst}) is that the weighted average, in regard to the compact kernel support, is computationally very demanding. As demonstrated by a convergence analysis by Zhu et al. \cite{Zhu_2015}, an accurate estimate of $V$ requires a small mean particle spacing $\Delta l$ compared to $h$ or $\Delta l \ll h$.
Nevertheless, in practice $h$ is chosen to be in the order of $\Delta l$ in order to limit the computational costs. By this pragmatic choice, volume partition errors are introduced which are incorporated in the momentum equation \cite{Hopkins_2015}. 
As a consequence, the actual Lagrangian particle trajectories are disturbed by numerical noise eventually causing pseudo-Lagrangian behaviour.
The latter triggers numerical dissipation \cite{Bauer_2012} and compromises the ability of WCSPH for an accurate simulation of vortex dynamics \cite{Hopkins_2015, Rossi_2015} and subsonic turbulence \cite{Bauer_2012, Hopkins_2015} at finite resolution. Although the noise is caused by numerical inaccuracy, within the scope of the paper it will be called thermal noise.\\
On the other hand, the pseudo-Lagrangian nature of the method is rooted in the deformation ability of the numerical particles. By imagining a cubic fluid element in a linear shear flow one would normally expect the fluid element to be deformed into a parallelepiped.
Conversely, a numerical particle will preserve its shape except for weak isotropic volume changes limited by the weakly-compressible character of the method. Hence, the particle suffers from a missing mechanism to adapt its shape according to the actual fluid flow. Vogelsberger et al. \cite{Vogelsberger_2012} even go one step further, arguing that at least the spherical kernel support associated to a center particle should be deformed in a shear flow. However, the shape of the kernel is practically forced to stay spherical which causes well-known mixing problems \cite{Vogelsberger_2012}. \\
Since the simulation of vortices, turbulence and mixing at finite resolution is problematic at the current stage due to the pseudo-Lagrangian character, we want to gain a deeper understanding of the latter. Therefore, a mathematical measure is inevitable. To obtain this measure, it is important to highlight the striking similarity of the estimate for $V$ in Eq. (\ref{eq:VolEst}) to the calculation of macroscopic quantities in non-equilibrium molecular dynamics (NEMD). There, converged macroscopic field quantities are obtained by an averaging process over sufficient molecules (particles) in a local compact domain (fluid element), as well (e.g. \cite{Hardy_1982, Smith_2015}). \\
Motivated by this observation, we take up the idea of Vogelsberger et al. \cite{Vogelsberger_2012} that the spherical kernel support can be interpreted as a fluid element. Based on that, we will develop an isothermal Navier-Stokes equation for a kernel-based fluid element by a NEMD analysis. The resulting momentum transport equation contains an additional term depending on a tensor field $\boldsymbol{\tau}_{mol}$ which we will call molecular stress tensor. It will be demonstrated that this term quantifies the influence of the previously described pseudo-Lagrangian character and gives direction to conceptual improvements of WCSPH. \\
So far, the motivation of this work was restricted to WCSPH as a weakly-compressible particle-based discretization method. However, the theory derived in the following will be applicable to all weakly-compressible particle-based discretization methods utilizing:

\begin{itemize}
\item A volume partitioning based on the approximation in Eq. (\ref{eq:VolEst}).

\item Particles as discretization volumes which are interconnected by local, spherical kernels.
\end{itemize}

\section{The isothermal kernel transport equations}
\label{sec:KernelTransport}

In this section, the isothermal transport equations for a kernel-based fluid element are derived by means of a NEMD analysis. We use the theory of Hardy \cite{Hardy_1982}, which is similar to the Irving and Kirkwood formalism \cite{Irving_1950}. The most important difference is in regard to the utilized localization function. Within the theory of Hardy \cite{Hardy_1982}, the Dirac delta function $\delta$ is replaced by a positive, symmetrical function with compact support. Obviously, the kernel function $W_h$ itself appears to be a natural choice. The procedure seems to be very promising since the theory of Hardy \cite{Hardy_1982} was already successfully deployed in the SPH context by Tartakovsky and Panchenko \cite{Tartakovsky_2016} to model surface tension forces. Furthermore, we believe that the derivation can be interpreted as a generalization of the work of Ellero et al. \cite{Ellero_2010}. In their work, the Irving and Kirkwood formalism was applied to a numerical SPH particle set to calculate implicit atomistic viscosities in a Couette flow. \\
For the first presentation of our new concept, we focus on the transport of mass and momentum and neglect the transport of entropy and chemical species. \\
In the following, quantities indexed with $j$ are results obtained from a numerical simulation by means of a weakly-compressible particle-based discretization method. \\
Starting point of the NEMD analysis is an estimate for the density $\rho$ at a fixed kernel center position $\mathbf{x} = const$. The position can coincide with a center particle but does not have to. The estimate includes Eq. (\ref{eq:VolEst}) for the volume: 
\begin{equation}
\rho = \frac{M}{V} = M \sum_{j=1}^{N_{NGB}} W_h(\mathbf{x}-\mathbf{x}_j)~.
\label{eq:Density}
\end{equation}
In the context of weakly-compressible particle-based discretization methods, Eq. (\ref{eq:VolEst}) is a commonly used approximation (e.g. \cite{Hopkins_2015, Frontiere_2017}). According to the theory of Hardy \cite{Hardy_1982}, we take the temporal derivative of Eq. (\ref{eq:Density}) at $\mathbf{x} = const $ and apply the chain rule. The resulting continuity equation for the kernel-based fluid element can be obtained by considering the particle positions $\mathbf{x}_j(t)$ to be only functions of time and $M=const$:
\begin{eqnarray}
\partial_t \rho &=& M \sum_{j=1}^{N_{NGB}} \partial_t W_h(\mathbf{x}-\mathbf{x}_j) \nonumber \\
&=& -M \sum_{j=1}^{N_{NGB}} \mathbf{v}_j \cdot \nabla W_h(\mathbf{x}-\mathbf{x}_j) \nonumber \\
&=& - \nabla \cdot \sum_{j=1}^{N_{NGB}} M \mathbf{v}_j W_h(\mathbf{x}-\mathbf{x}_j) = - \nabla \cdot \mathbf{p} ~,
\label{eq:Continuity}
\end{eqnarray}
where $\partial_t \mathbf{x}_j = \mathbf{v}_j \in \mathbb{R}^3$ denotes the particle velocities. The derived continuity equation  Eq. (\ref{eq:Continuity}) automatically leads to a natural definition of the kernel associated momentum density $\mathbf{p} \in \mathbb{R}^3$ within the particle framework. Since the latter can be analytically expressed by $\mathbf{p} = \rho \mathbf{U}$, $\mathbf{U}$ denoting the kernel velocity, a comparison with Eq. (\ref{eq:Density}) results in a definition of $\mathbf{U}$:
\begin{eqnarray}
\mathbf{p} = \rho \mathbf{U} \stackrel{!}{=} \sum_{j=1}^{N_{NGB}} M \mathbf{v}_j W_h(\mathbf{x}-\mathbf{x}_j) \nonumber \\ 
\Rightarrow \mathbf{U} := \frac{\sum_{j=1}^{N_{NGB}} \mathbf{v}_j W_h(\mathbf{x}-\mathbf{x}_j)}{\sum_{j=1}^{N_{NGB}} W_h(\mathbf{x}-\mathbf{x}_j)} ~.
\label{eq:KernelVelocity}
\end{eqnarray}
The focus on weakly-compressible flows ($V \approx const$) allows us to conclude that the kernel velocity $\mathbf{U}$ in Eq. (\ref{eq:KernelVelocity}) is similar to a Shepard filtered velocity \cite{Vignjevic_2000}. The Shepard filter is often employed in the SPH community to restore zero-order consistency (e.g. \cite{Vignjevic_2000, Lastiwka_2005, Colagrossi_2009}). Physically speaking, the thermal noise introduced by Eq. (\ref{eq:VolEst}) is partially cancelled out at the kernel level in Eq. (\ref{eq:KernelVelocity}). Hence, in correspondence to NEMD, it justifies a separation of the particle velocities $\mathbf{v}_j$ into the local kernel velocity $\mathbf{U}$ and the peculiar velocities $\mathbf{w}_j$:
\begin{equation}
\mathbf{v}_j = \mathbf{U} + \mathbf{w}_j~.
\label{eq:VDecomp}
\end{equation}
Analogously to NEMD, the peculiar velocity $\mathbf{w}_j$ contains not only the thermal noise introduced by Eq. (\ref{eq:VolEst}) but also the information of the velocity field within the kernel. Even in a perfectly approximated flow where the discretized velocities $\mathbf{v}_j$ match the corresponding analytical field quantities, the mere existence of a kernel introduces a non-vanishing peculiar velocity $\mathbf{w}_j$. This thought will be addressed in section \ref{sec:Convergence} in more detail. \\
If we repeat the procedure applied to the density $\rho$ in Eq. (\ref{eq:Continuity}) with the momentum density $\mathbf{p}$, we end up with a momentum transfer equation for the kernel-based fluid element. Therefore, we calculate the temporal derivative of Eq. (\ref{eq:KernelVelocity}). Taking into account that the particle velocities $\mathbf{v}_j(t)$ are only functions of time as well as Eq. (\ref{eq:VDecomp}), it follows:
\begin{widetext}
\begin{eqnarray}
\partial_t \mathbf{p} &=& M \sum_{j=1}^{N_{NGB}}  \mathbf{v}_j \partial_t W_h(\mathbf{x}-\mathbf{x}_j) + \mathbf{a}_jW_h(\mathbf{x}-\mathbf{x}_j) = -M \sum_{j=1}^{N_{NGB}}  \mathbf{v}_j \mathbf{v}_j^T \nabla W_h(\mathbf{x}-\mathbf{x}_j) + M \sum_{j=1}^{N_{NGB}} \mathbf{a}_jW_h(\mathbf{x}-\mathbf{x}_j) \nonumber \\
&=& div \left( \sum_{j=1}^{N_{NGB}} -M \mathbf{v}_j \mathbf{v}_j^T W_h(\mathbf{x}-\mathbf{x}_j) \right) + M \sum_{j=1}^{N_{NGB}} \mathbf{a}_jW_h(\mathbf{x}-\mathbf{x}_j) \nonumber \\
&=& div \left( \sum_{j=1}^{N_{NGB}} -M (\mathbf{U}+\mathbf{w}_j)(\mathbf{U}+\mathbf{w}_j)^T W_h(\mathbf{x}-\mathbf{x}_j) \right) + M \sum_{j=1}^{N_{NGB}} \mathbf{a}_jW_h(\mathbf{x}-\mathbf{x}_j) ~,
\label{eq:MomTransfer}
\end{eqnarray}
\end{widetext}
where the superscript $T$ denotes a transposed vector field and $\partial_t \mathbf{v}_j=\mathbf{a}_j$ the particle accelerations. The operator $div$ represents the divergence acting on a tensor field. If we further consider Eq. (\ref{eq:Density}) in the last line of Eq. (\ref{eq:MomTransfer}) and the fact that the kernel associated sum of the peculiar momenta vanish per definition of Eq. (\ref{eq:VDecomp}) (Appendix \ref{sec:Appendix1}), namely
\begin{equation}
\sum_{j=1}^{N_{NGB}} M \mathbf{w}_j W_h(\mathbf{x}-\mathbf{x}_j)=0~,
\label{eq:PeculiarMomenta}
\end{equation}
the mixed terms with $\mathbf{U}\mathbf{w}_j^T$ and $\mathbf{w}_j\mathbf{U}^T$ cancel out and we obtain:
\begin{eqnarray}
\partial_t \mathbf{p} &=& div \left( \sum_{j=1}^{N_{NGB}} -M \mathbf{w}_j\mathbf{w}_j^T W_h(\mathbf{x}-\mathbf{x}_j) \right) \nonumber \\
&+& M \sum_{j=1}^{N_{NGB}} \mathbf{a}_jW_h(\mathbf{x}-\mathbf{x}_j) - div(\rho \mathbf{U}\mathbf{U}^T)~. 
\label{eq:MomTransfer2}
\end{eqnarray}
The final form of the kernel momentum transfer equation is derived by applying the following steps: First, we rearrange the last term on the right hand side (rhs) to the left hand side (lhs) and consider Eq. (\ref{eq:Continuity}). Second, we introduce the Lagrangian derivative  $\frac{d}{dt} = \partial_t + \mathbf{U} \cdot \nabla$ on the kernel level. Third, we argue that $V \approx V_j$, due to the weakly-compressible assumption. If we finally abbreviate the sum in the first term of Eq. (\ref{eq:MomTransfer2}) on the rhs with $\boldsymbol{\tau}_{mol}$, the momentum transfer equation reads:
\begin{equation}
\rho \frac{d\mathbf{U}}{dt} = div( \boldsymbol{\tau}_{mol}) + \rho \sum_{j=1}^{N_{NGB}} \mathbf{a}_jW_h(\mathbf{x}-\mathbf{x}_j) V_j~.
\label{eq:KernelNavierStokes}
\end{equation}
Obviously, Eq. (\ref{eq:KernelNavierStokes}) links the forces acting on the kernel-based fluid element (lhs) to the numerically approximated particle forces, i.e. $\mathbf{a}_j$ (second term on the rhs). \\
In the next section we will demonstrate that the set of Eq. (\ref{eq:Continuity}) and Eq. (\ref{eq:KernelNavierStokes}), subsequently called \emph{isothermal kernel transport equations}, reveal a physical interpretation of mathematical convergence in weakly-compressible particle-based discretization methods. In particular the derived molecular stress tensor 
\begin{equation}
\boldsymbol{\tau}_{mol} := \sum_{j=1}^{N_{NGB}} -M \mathbf{w}_j\mathbf{w}_j^T W_h(\mathbf{x}-\mathbf{x}_j)
\label{eq:MolecularStressTensor}
\end{equation}
will be of paramount importance. The latter proves to be a mathematical measure for quantifying the pseudo-Lagrangian behaviour of weakly-compressible particle based discretization methods.

\section{A physical interpretation of mathematical convergence}
\label{sec:Convergence}

Although the isothermal kernel transport equations are not directly considered in  weakly-compressible particle simulations, they can be studied theoretically. This is the objective of this section. As a result, a new, physically comprehensible interpretation of convergence is introduced, as well as an interpretation for $\boldsymbol{\tau}_{mol}$ in Eq. (\ref{eq:MolecularStressTensor}). \\
We assume a convergent weakly-compressible particle-based discretization method. Then, in the formal limit 
\begin{equation}
W_h \rightarrow \delta	\quad \& \quad \sum \rightarrow \int~,
\label{eq:limit}
\end{equation}
the isothermal fluid dynamic transport equations should be reproduced exactly by the discretized equations. In this case 
\begin{enumerate}
\item The particle quantities should match the analytical field quantities of the actual fluid elements, namely
\begin{equation}
\mathbf{v}_j \rightarrow \mathbf{v}(\mathbf{x},t)~, \quad \mathbf{a}_j \rightarrow \mathbf{a}(\mathbf{x},t)~.
\end{equation}
\item The kernel summation should reproduce an exact partition of unity
\begin{equation}
\sum_{j=1}^{N_{NGB}} W_h(\mathbf{x}-\mathbf{x}_j) V_j \rightarrow \int \delta ~d\mathbf{x}=1 ~.
\end{equation}
\end{enumerate}
Thus, the kernel density in Eq. (\ref{eq:Density}) and the kernel velocity in Eq. (\ref{eq:KernelVelocity}) coincide with their analytical counterparts as well, i.e. $\rho = \rho(\mathbf{x},t)$ and $\mathbf{U}=\mathbf{v}(\mathbf{x},t)$. Hence, Eq. (\ref{eq:Continuity}) converges to the actual continuity equation. Physically, the smoothing procedure, or more specifically low-pass filtering, introduced as approximation, is eliminated. \\
The same expected result can be obtained for the second term on the rhs of the kernel momentum transfer equation Eq. (\ref{eq:KernelNavierStokes}). Again, in the limit (\ref{eq:limit}), the smoothing is eliminated and the volumetric forces $\rho \mathbf{a}(\mathbf{x},t)$ remain. New insights come from the analysis of the first term on the rhs in Eq. (\ref{eq:KernelNavierStokes}). Obviously, this term is cancelled out because the peculiar velocities in Eq. (\ref{eq:VDecomp}) become zero. \\
Although this finally illustrates that Eq. (\ref{eq:KernelNavierStokes}) converges to the actual momentum transfer equation as required, it indeed brings up a question. What is an appropriate physical interpretation of the first term on the rhs of Eq. (\ref{eq:KernelNavierStokes}) and especially the molecular stress tensor $\boldsymbol{\tau}_{mol}$ in Eq. (\ref{eq:MolecularStressTensor})?\\
\begin{figure}[t]
\centering
\includegraphics[trim= 0cm 0cm 0cm 0cm, width=3in]{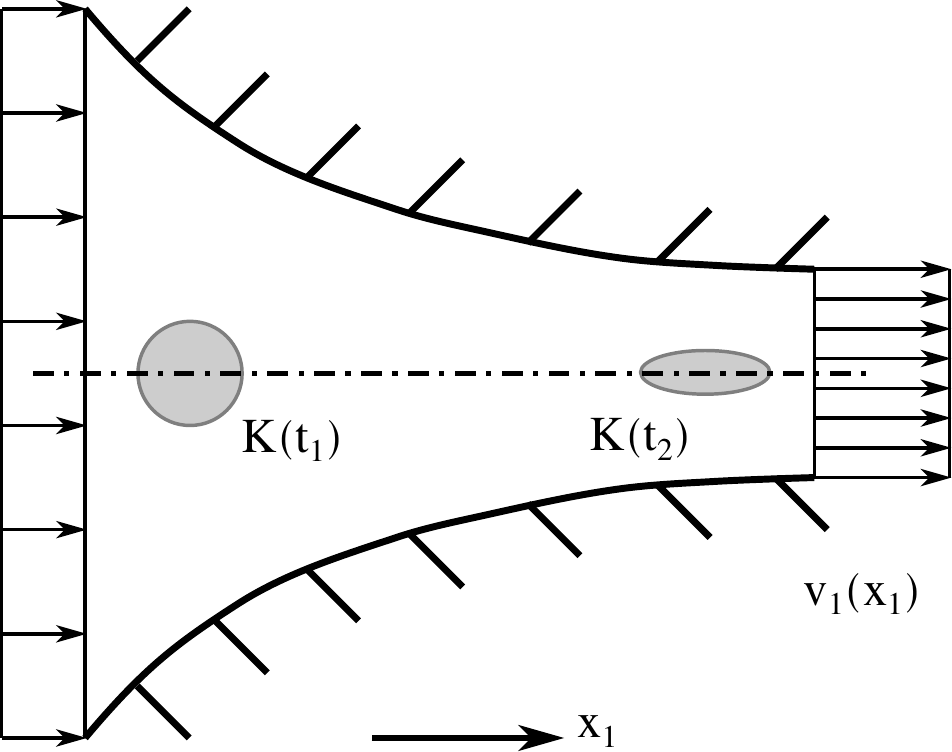}
\caption{Generic one-dimensional, axisymmetric nozzle flow considered to demonstrate the effect of the molecular stress tensor on a kernel-based fluid element $K$. The element at two different times $t_1$ and $t_2$ along its Lagrangian flow path is depicted as a grey patch.}
\label{fig:GenericFlow}
\end{figure}
In NEMD the term $div(\boldsymbol{\tau}_{mol})$ is called the kinetic contribution and arises from the time derivative of the localization function used to define a fluid element (see Eq. (\ref{eq:MomTransfer})) \cite{Hardy_1982}. The symmetric tensor $\boldsymbol{\tau}_{mol}$ is a quadratic form in the peculiar velocity $\mathbf{w}_j$ and possesses the dimension of a stress. A dimensional analysis gives a quick proof. Since the peculiar velocities $\mathbf{w}_j$ describe a velocity relative to the averaged velocity $\mathbf{U}$ (see Eq. (\ref{eq:KernelVelocity}) and (\ref{eq:VDecomp})), the components of $\boldsymbol{\tau}_{mol}$ are only zero in a perfectly homogeneous flow. Noisy and non-homogeneous flows are always exposed to a molecular stress due to the mere presence of a kernel. In this case, it is obvious that the numerically approximated flow will be implicitly affected in its evolution. \\
Further analysis of the kinetic effect of spatially varying molecular stresses ($div(\boldsymbol{\tau}_{mol})$ in Eq. (\ref{eq:KernelNavierStokes})) on a kernel-based fluid element $K$, will be performed using a simple and generic flow field. We consider the incompressible, stationary flow through an axisymmetric nozzle as depicted in Fig. (\ref{fig:GenericFlow}). Additionally, the nozzle is assumed to be very short. Hence, the boundary layer has not enough time to develop such that the velocity profile along the coordinate $x_1$ can always be approximated locally as a plug flow. Consequently, the flow problem is one-dimensional. Given the assumptions above, the volumetric flow rate in the nozzle is constant. There is only the velocity component $v_1$ in $x_1$ direction which is inversely proportional to the local cross-sectional area $A$, namely $v_1 \sim A^{-1}$. Since we are only interested in the nozzle geometry, we choose $A \sim x^{-2}$ as a parametrization of the cross-sectional area and obtain:
\begin{equation}
v_1(x_1) \sim x_1^2~.
\label{eq:VelProfil}
\end{equation}
For the calculation of the molecular stress tensor $\boldsymbol{\tau}_{mol}$ in Eq. (\ref{eq:MolecularStressTensor}) it is indispensable to discretize the problem. Therefore, the fluid domain as depicted in Fig. (\ref{fig:GenericFlow}) is decomposed into finite mass particles arranged on a cartesian lattice. The cartesian arrangement was chosen due to simplicity but can be replaced by an arbitrary particle distribution. To each particle the velocity given by Eq. (\ref{eq:VelProfil}) is assigned which implies that the influence of noise is neglected. Furthermore, a spherical kernel-based fluid element $K$ is introduced which is highlighted in Fig. (\ref{fig:GenericFlow}) as grey patch $K(t_1)$.

\begin{figure}[ht]
\centering
\includegraphics[trim= 0cm 0cm 0cm 0cm, width=2in]{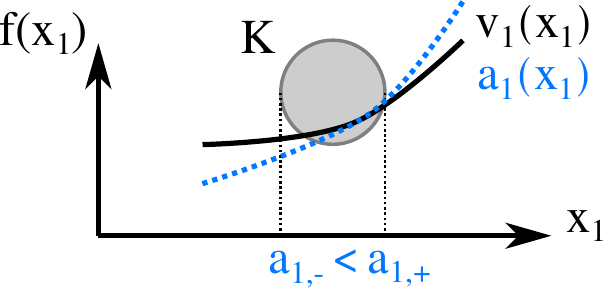}
\caption{Kernel-based fluid element $K$ exposed to a quadratically increasing velocity. The corresponding acceleration is highlighted as blue dashed line.}
\label{fig:KernelStretching}
\end{figure}
Before we start to demonstrate the impact of $div(\boldsymbol{\tau}_{mol})$ on $K$, it is important to understand how $K$ would normally deform along its Lagrangian flow path. Therefore, the knowledge of the forces or accelerations acting on the boundaries of $K$ is necessary. A kernel-based fluid element $K$ exposed to a quadratic flow field is depicted in Fig. (\ref{fig:KernelStretching}). The Lagrangian acceleration is given by $a_1(x_1) = v_1\frac{dv_1}{dx_1} \sim x_1^{3}$ and is illustrated as blue dashed line in Fig. (\ref{fig:KernelStretching}). Consequently, $K$ is exposed to unequal forces on the left ($-$) and right ($+$) boundary and in the situation considered $a_{1,-} < a_{1,+}$ holds. This imbalance implies that the element $K$ will be stretched along its flow path as indicated in Fig. (\ref{fig:GenericFlow}). Hence, an initially spherical element $K(t_1)$ is deformed into a stretched ellipsoid $K(t_2)$ at a later time $t_2$. \\
Unfortunately, the deformation behaviour of $K$ is neglected in weakly-compressible particle based discretization methods. The pseudo-Lagrangian character forces the kernel to stay spherical. As it will be demonstrated subsequently, the tensor field $\boldsymbol{\tau}_{mol}$ in Eq. (\ref{eq:MolecularStressTensor})  prevents the actual deformation by the introduction of local molecular stresses. The only non-zero component of the molecular stress tensor regarding the previous assumptions is:
\begin{equation}
\tau_{mol,11} = \sum_{j=1}^{N_{NGB}} -M w_{j,1}^2  W_h(\mathbf{x}-\mathbf{x}_j) < 0~.
\label{eq:Mol11}
\end{equation}
Since each factor in the summation is obviously positive, the component $\tau_{mol,11}$ will be negative in the whole domain. If we now consider $\mathbf{n}$ as a vector pointing into the main flow direction, the resulting stress vector $\mathbf{t}=\boldsymbol{\tau}_{mol}\mathbf{n}$ will be antiparallel to the main flow. From this observation one may conclude that the molecular stresses introduce additional friction on the particle level and decelerate the flow. As can be seen from Eq. (\ref{eq:KernelNavierStokes}), this is only true if  $\partial_1\tau_{mol,11} < 0$ or, in other words, the stresses are growing downstream. Hence, for the chosen example, it is inevitable to understand how $w_{j,1}^2$ change in $x_1$ direction. As the peculiar velocity component $w_{j,1}$ describes a velocity relative to the averaged velocity $U_1$ (Eq. (\ref{eq:VDecomp})), the distribution of $w_{j,1}$ inside a kernel-based element $K$ downstream of the nozzle can only change if the velocity field has a non-zero curvature. This is demonstrated in Appendix \ref{sec:Appendix2}. From Eq. (\ref{eq:VelProfil}), we conclude that the resulting curvature in our example is $\frac{d^2v_1}{dx_1^2}=2~1/(ms)$ and positive everywhere. Considering that, it is obvious that $w_{j,1}^2$ is growing in the main flow direction and that $\partial_1\tau_{mol,11} < 0$ holds. \\
The example demonstrates that spatially changing molecular stresses, as described by $div(\boldsymbol{\tau}_{mol})$ in Eq. (\ref{eq:KernelNavierStokes}), will adjust the flow field in such a way that kernel-based fluid elements $K$ will preserve their spherical shape. In the given example, the flow is thus decelerated. Since the momentum transfer is driven by the curvature of the velocity field, it can be concluded that $div(\boldsymbol{\tau}_{mol})$ describes an additional diffusive momentum transfer. The latter relation is well-known from general transport phenomena  \cite{Baehr_2011} and is causing a homogenization of the driving field quantity. \\
Finally, we draw the following conclusions: Up to now, mathematical convergence of weakly-compressible particle based discretization schemes could be physically interpreted as the elimination of a smoothing procedure or more specifically low-pass filtering. We could demonstrate in this section, by deriving the kernel transport equations in section \ref{sec:KernelTransport}, that mathematical convergence also means that the pseudo-Lagrangian character, quantified by the molecular stresses $\boldsymbol{\tau}_{mol}$ in Eq. (\ref{eq:MolecularStressTensor}), is eliminated. The stresses are caused by non-zero peculiar velocities $\mathbf{w}_j$, which result from the introduction of a spherical, finite size kernel in a noisy approximated, inhomogeneous flow field. As $\mathbf{w}_j$ is quadratically fed back into $\boldsymbol{\tau}_{mol}$ and $\mathbf{w}_j$ is bigger in velocity fields with high local curvature, we hypothesize that these implicit molecular stresses could be the reason why weakly-compressible particle based discretization schemes show inferior performance in strongly subsonic turbulent flows \cite{Zhu_2015, Hopkins_2015, Bauer_2012}. \\
So far, the current study has been focused on the derivation of a mathematical background to quantify the pseudo-Lagrangian character and the consecutive, theoretical interpretation of the derived molecular stresses $\boldsymbol{\tau}_{mol}$. The objective of the next section \ref{sec:IndicatorAndConcept} is to demonstrate that $\boldsymbol{\tau}_{mol}$ can also be used as a quality indicator in simulations.

\section{Towards Conceptual Improvements: Molecular Stresses as a Quality Indicator}
\label{sec:IndicatorAndConcept}

As discussed previously, accurate simulations of vortices, turbulent flows and mixing processes are an issue for weakly-compressible particle based discretization methods in general \cite{Vogelsberger_2012, Zhu_2015, Hopkins_2015, Bauer_2012, Rossi_2015, Deng_2019}. We will demonstrate the practical importance of $\boldsymbol{\tau}_{mol}$ as a quality indicator for a two-dimensional turbulent flow. We have tested our concept also with laminar flows and the conclusions are identical.

\subsection{Description of the numerical setup}

\begin{table*}[t]
%%% increase table row spacing, adjust to taste
\renewcommand{\arraystretch}{1.3}
\caption{Parameters utilized for the investigated Kolmogorov flow.}
\label{tab:Parameters}
\begin{ruledtabular}
%%% Some packages, such as MDW tools, offer better commands for making tables
%%% than the plain LaTeX2e tabular which is used here.
\begin{tabular}{ccccccccc}
 $\rho_{ref}$ [$kg/m\textsuperscript{3}$] & $p_{ref}$ [$Pa$] & $c_s$ [$m/s$] & $\nu$ [$m\textsuperscript{2}/s$] & $\chi$ [$m/s\textsuperscript{2}$] &  $k_2$ [$1/m$] & $\mu$ [$1/s$] & $L$ [$m$] & $\Delta l$ [$\mu m$] \\
\hline
 1000 & 400 & 2 & 1.6 $\cdot$ 10\textsuperscript{-6} & 1.75 & 1047.2 & 0.7 & 0.042 & 125\\
\end{tabular}
\end{ruledtabular}
\end{table*}

For the subsequent numerical tests, the experiment conducted by Rivera and Wu \cite{Rivera_2000} is considered. In their study forced, statistically steady two-dimensional turbulence was created in a freely suspended soap film driven by electromagnetical forcing. As the authors point out, the resulting fluid flow can be described by an incompressible Kolmogorov flow \cite{Zhang_2019} represented by an augmented Navier-Stokes equation:
\begin{eqnarray}
& \frac{1}{\rho}\frac{d\rho}{dt} = - \nabla \cdot \mathbf{v} \Rightarrow \nabla \cdot \mathbf{v} = 0~, \nonumber \\
& \frac{d\mathbf{v}}{dt} = -\frac{\nabla p}{\rho} + \nu \Delta \mathbf{v} + \chi sin(k_2 x_2) \mathbf{e}_1 - \mu \mathbf{v}~. 
\label{eq:KolmogorovFlow}
\end{eqnarray}
In Eq. (\ref{eq:KolmogorovFlow}) the symbol $\mathbf{v} \in \mathbb{R}^2$ describes the two-dimensional velocity field, $p$ the pressure field and $\nu$ the kinematic viscosity. The third term on the rhs of Eq. (\ref{eq:KolmogorovFlow}) models a sinusoidal forcing with amplitude $\chi$ and wavenumber $k_2$ pointing into the direction of the $x_1$ coordinate given by the vector $\mathbf{e}_1 = (1, 0)^T$. As two-dimensional turbulence is characterized by an inverse cascade process, a large scale friction, namely $\mu \mathbf{v}$ in Eq. (\ref{eq:KolmogorovFlow}), has to be introduced to ensure a statistically steady flow \cite{Bofetta_2012}. The latter is known as Rayleigh drag \cite{Rivera_2000} and its magnitude is determined by the friction coefficient $\mu$. \\
To demonstrate that even sophisticated, higher order weakly-compressible particle based discretization methods suffer from the pseudo-Lagrangian character, we have solved Eq. (\ref{eq:KolmogorovFlow}) numerically with the Meshless-Finite-Mass method (MFM) as introduced by Hopkins \cite{Hopkins_2015}. For this reason, we have augmented the open-source code GIZMO \cite{Hopkins_2017} by the sinusoidal forcing and the Rayleigh drag as given by Eq. (\ref{eq:KolmogorovFlow}), which is a straightforward task. Additionally, we have implemented the Cole equation of state \cite{Cole_1948} defined by:
\begin{equation}
p = p_{ref} + \rho_{ref}c_s^2\left(\frac{\rho}{\rho_{ref}} - 1\right)~.
\label{eq:Cole}
\end{equation}
In the WCSPH literature, Eq. (\ref{eq:Cole}) is well-known (e.g. \cite{Monoghan_1994, Marrone_2013}) and utilized to close the set of partial differential equations in Eq. (\ref{eq:KolmogorovFlow}). The parameters introduced in Eq. (\ref{eq:Cole}) are a constant reference density $\rho_{ref}$, an artificial speed of sound $c_s$ and the constant background pressure $p_{ref}$. Although a choice of $p_{ref}>0$ enhances the numerical dissipation by the generation of local high frequency modes \cite{Hopkins_2015, Marrone_2013, Colagrossi_2012}, in the methods of interest imposing a background pressure $p_{ref}>0$ is essential  because it stabilizes the simulation in spatial regions of strong deformation by implicit particle regularization \cite{Marrone_2013}. \\
The numerical domain is defined as a periodic, square box of length $L=0.042~m$ which corresponds to exactly seven wavelengths of the sinusoidal forcing in Eq. (\ref{eq:KolmogorovFlow}) and is in accordance with Fig. (1) of \cite{Rivera_2000}. The finite mass particles are initially at rest and placed on an equidistant cartesian grid with a spacing of $\Delta l = 125~\mu m$. Since the viscous friction scale in the laboratory experiment was determined to be $l_{\nu} = 250~\mu m$, under the assumption that a particle represents a fluid element, we would conduct a resolved direct numerical simulation. As kernel the Wendland C4 kernel was utilized \cite{Dehnen_2012} with an effective number of neighbors $N_{NGB} =32 $ \cite{Hopkins_2015}. The numerical parameters in Eq. (\ref{eq:KolmogorovFlow}) and Eq. (\ref{eq:Cole}) as used for the simulation are listed in TABLE \ref{tab:Parameters}. They are similar to the ones documented in \cite{Rivera_2000, Rivera_2001}. Only $\rho_{ref}$, $p_{ref}$ and $\chi$ had to be chosen as they were not reported. Since the soap film is realized with an aqueous solution \cite{Rivera_2000} it seems reasonable to set $\rho_{ref} = 1000~kg/m^3$. The background pressure was set to the smallest possible choice $p_{ref}=400~Pa$ with the goal of stabilizing the simulation on the one hand and minimizing the numerical dissipation on the other hand. The forcing amplitude $\chi$ was determined from a parametric study with the goal to reach the same time-averaged root-mean-square velocity 
\begin{equation}
u_{rms} = \frac{1}{|I_T|} \int_{I_T} \sqrt{\frac{1}{A}\int_A \mathbf{v}^2 d\mathbf{x}}~dt \approx 0.11~m/s
\label{eq:Calibration}
\end{equation} 
as in the publication of Rivera \cite{Rivera_2000}. The outcome of the parameter study was that for $\chi = 1.75~m/s^2$ the value for $u_{rms}$ could be reproduced. Hence, the time averaged level of turbulent kinetic energy in the system was matched, but, as demonstrated in the next subsection, only by an excessive energy input into the system which can be attributed to the pseudo-Lagrangian character. \\
The time range covered by the simulation was $T_{sim} = 8.4 ~s$.

\subsection{Analysis of the results}

\begin{figure*}[ht]

\begin{subfigure}{.49\textwidth}
\centering
\includegraphics[clip, trim= 0.5cm 1.5cm 0.5cm 2.2cm, width=3.1in]{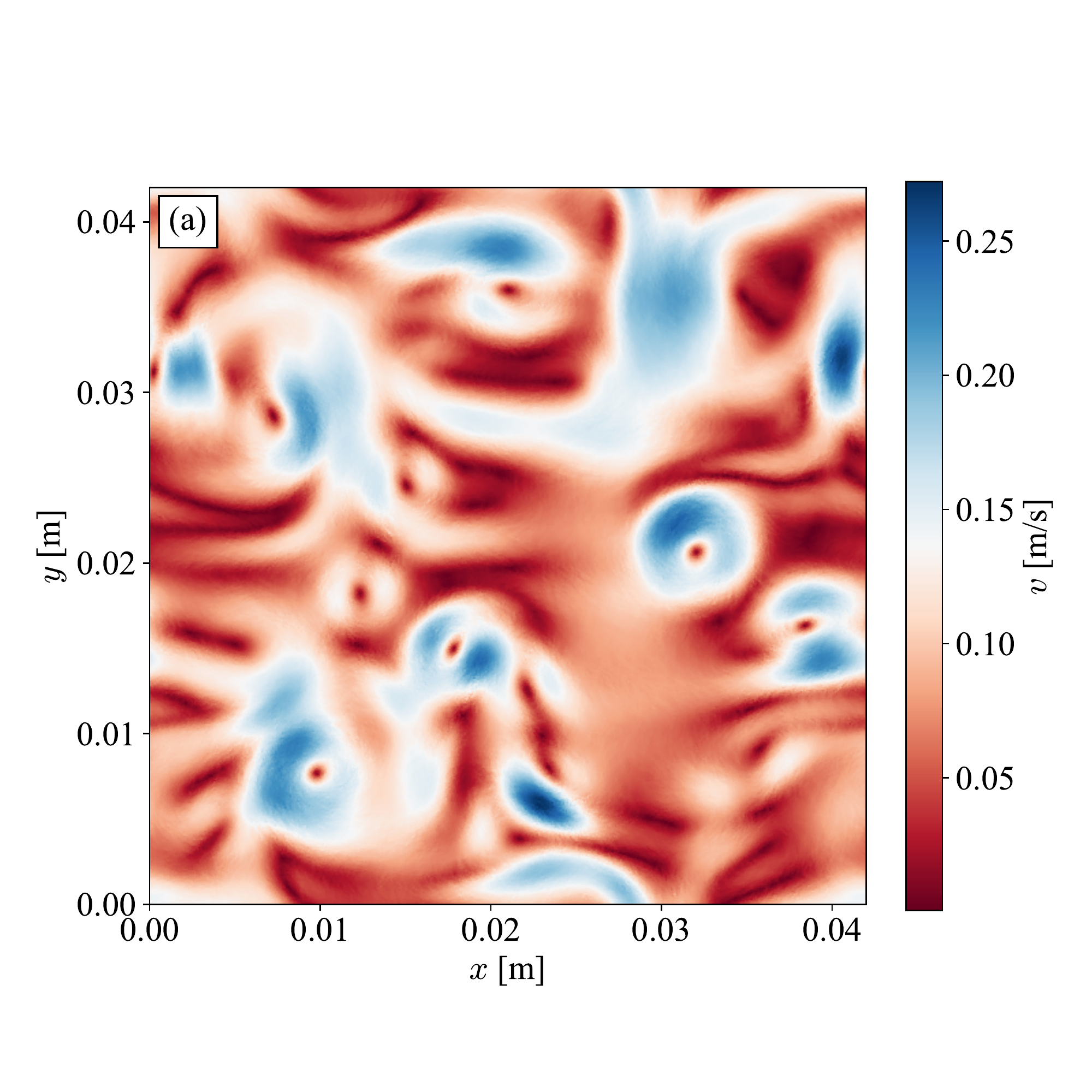}
%\caption{Snapshot of the velocity magnitude.}
\label{fig:SnapShotA}
\end{subfigure}
\hfill
\begin{subfigure}{.49\textwidth}
\centering
\includegraphics[clip, trim= 0.5cm 1.5cm 0.5cm 2.2cm, width=3.1in]{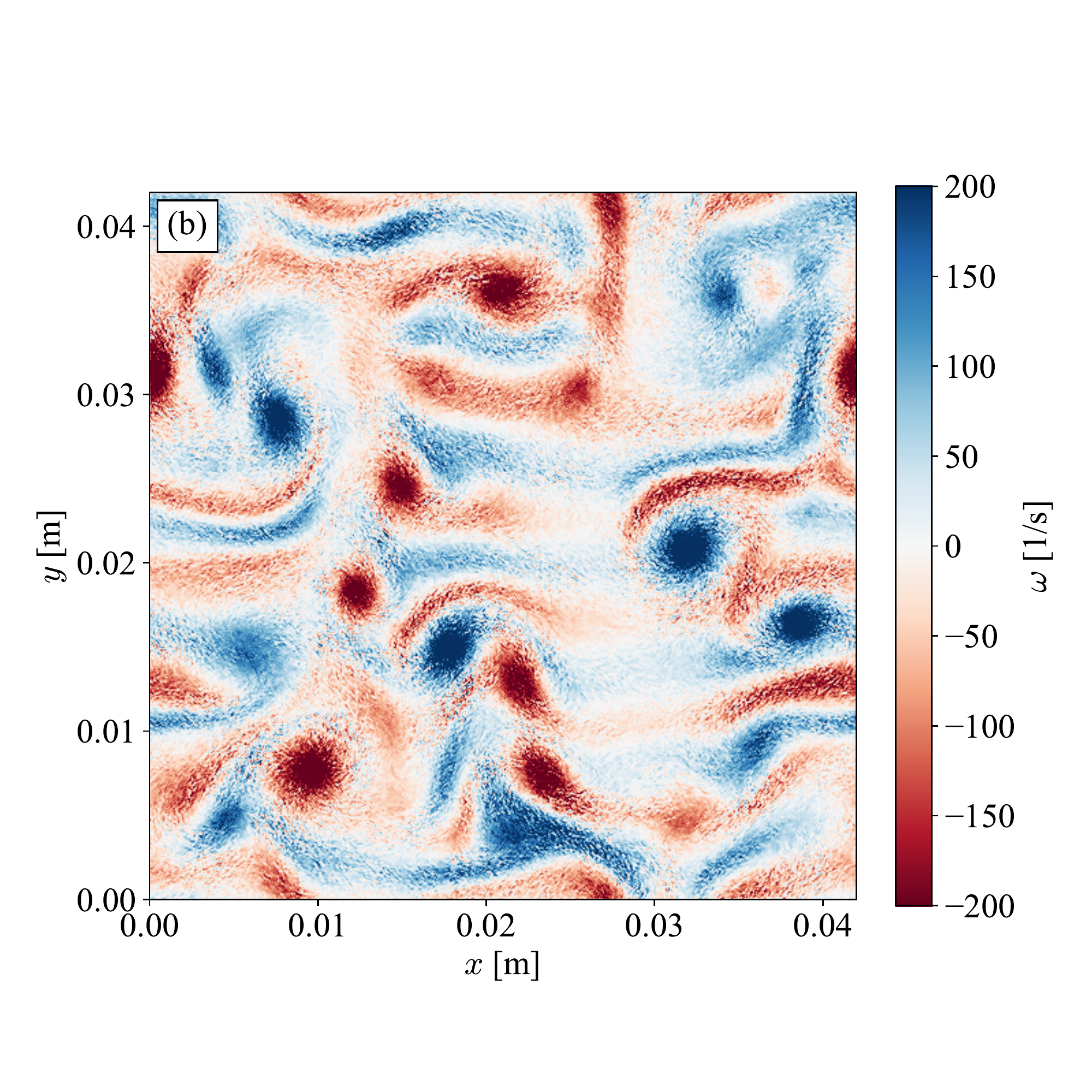}
%\caption{Snapshot of the vorticity field.}
\label{fig:SnapShotB}
\end{subfigure}
 
\caption{Snapshots of the statistically steady turbulent flow field at time $t^*=4.1~s$. (a) Snapshot of the velocity magnitude. (b) Snapshot of the vorticity field.}
\label{fig:SnapShots}
\end{figure*}

\begin{figure}[b]
\centering
\includegraphics[clip, trim= 0cm 0cm 0cm 0.5cm, width=3.1in]{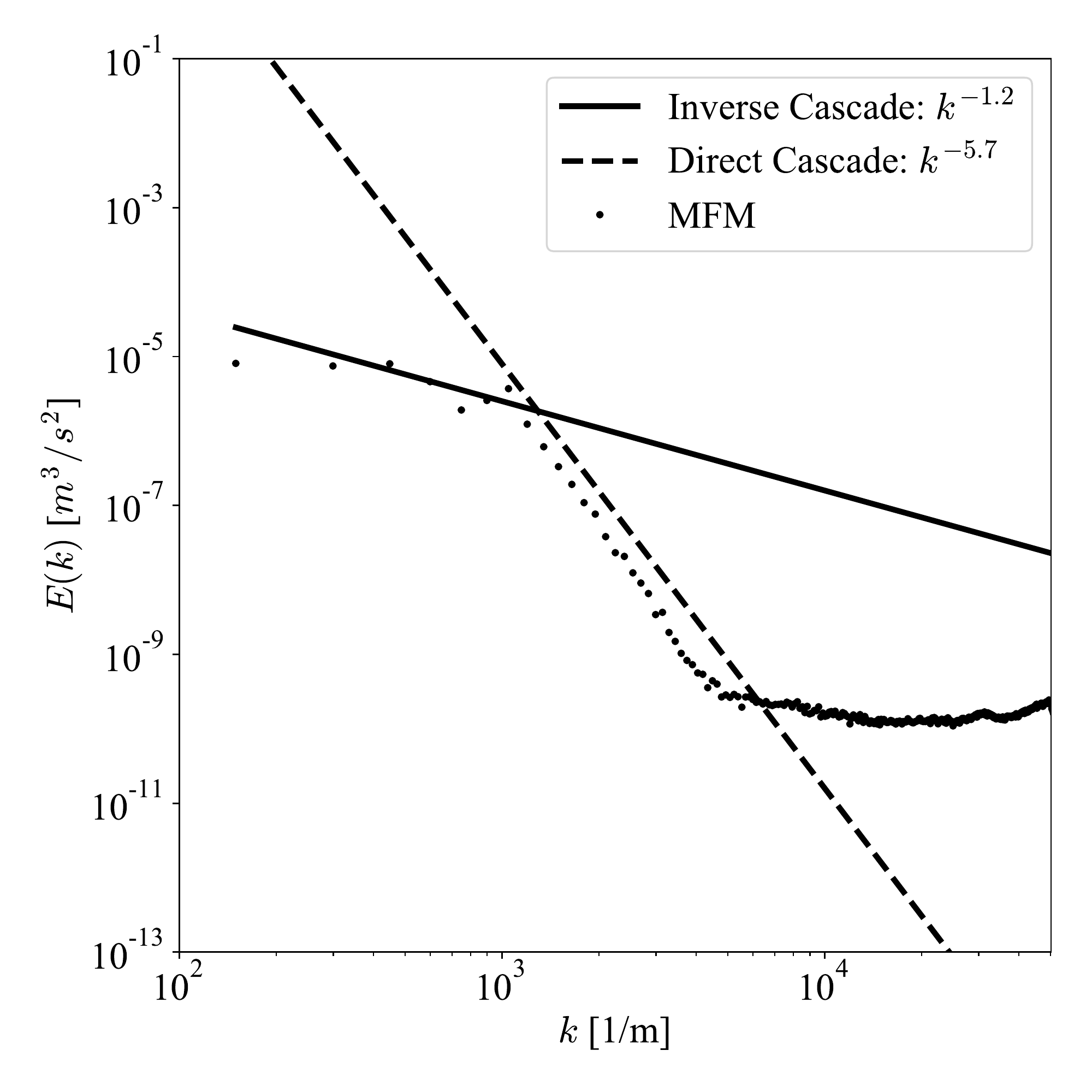}
\caption{Eulerian turbulent kinetic energy spectrum at time $t^*=4.1~s$.}
\label{fig:TurbulentSpectrum}
\end{figure}

After an initial transient process of $\Delta t_1 \approx 3~s$ in which the particles where accelerated due to the forcing in Eq. (\ref{eq:KolmogorovFlow}) and the turbulent flow develops, the flow reaches a statistically steady state as expected. In Fig. (\ref{fig:SnapShots}) snapshots of the stationary turbulence, namely of the velocity magnitude field $v$ and the vorticity field $\omega$, are depicted at a time $t^*=4.1~s$.
The vorticity $\omega = \partial_1 v_2 - \partial_2 v_1$ was extracted in a post-processing step by nearest neighbor sampling of the velocity field on a cartesian grid twice as fine as $\Delta l$ and consecutive application of a second order finite-difference method for the derivative approximation. Due to this sampling resolution, aliasing effects could be avoided \cite{Bauer_2012}.
\\
As it can be seen in the velocity field in Fig. (\ref{fig:SnapShots}a), the flow field contains complex flow features characterized by different spatial scales. Obviously, regions of strong shear as well as vortices are apparent. These features can be identified by regions where the color changes quickly over a short distance and by small red spots trapped by blue, high velocity areas. This becomes even more evident in the corresponding vorticity field $\omega$ in Fig. (\ref{fig:SnapShots}b). Here, regions of strong shear are highlighted by characteristic high vorticity filaments and vortices by circular, high vorticity spots. A comparison with the equivalent experimental field quantities in \cite{Rivera_2000} reveals similar results. \\
Furthermore, we have extracted the Eulerian turbulent kinetic energy spectrum at the time $t^* = 4.1 ~s$ by the utilization of the interpolated velocity field already used for the calculation of the vorticity field and the method described by Durran et al. \cite{Durran_2017}. The result is depicted in Fig. \ref{fig:TurbulentSpectrum} and clearly demonstrates that the spectral characteristics of two-dimensional turbulence are captured by the method. Similar to the work of Rivera and Ecke \cite{Rivera_2016}, the spectral energy of the inverse cascade scales with $E_{inv} \sim k^{-1.2}$ and the spectral energy of the direct cascade scales with $E_{dir} \sim k^{-5.7}$. It is interesting to note that for wavenumbers $k \gtrapprox 6000~1/m$, corresponding to the order of the kernel diameter for $N_{NGB} = 32$, the level of the spectrum saturates. Effectively, the 
direct cascade is inhibited by this kernel bottleneck which we think is related to the pseudo-Lagrangian character. \\
Although a typical two-dimensional turbulent flow field with the same amount of turbulent kinetic energy (see Eq. (\ref{eq:Calibration})) has developed, it can be easily demonstrated that this could only be achieved by an excessive energy input. According to Eq. (5) in the work of Rivera and Wu \cite{Rivera_2000}, the averaged, stationary energy balance for the turbulent kinetic energy is given by:
\begin{equation}
\epsilon_{inj} = \epsilon_{\nu} + \epsilon_{Rayleigh} = \nu\Omega + \mu u_{rms}^2~.
\label{eq:EnergyBalance}
\end{equation}
In Eq. (\ref{eq:EnergyBalance}), the averaged energy input $\epsilon_{inj}$ due to the forcing in Eq. (\ref{eq:KolmogorovFlow}) must balance the averaged frictional losses $\epsilon_{\nu}$, $\epsilon_{Rayleigh}$ on the rhs to ensure stationary turbulence. Since $u_{rms}$ was calibrated according to Eq. (\ref{eq:Calibration}), the loss $\epsilon_{Rayleigh} = \mu u_{rms}^2 $ due to Rayleigh drag is identical to the laboratory experiment. Hence, the averaged level of viscous dissipation $\epsilon_{\nu}$ determines the amount of injected turbulent kinetic energy. For an incompressible flow the viscous dissipation is directly proportional to the averaged enstrophy $\Omega$ \cite{Colagrossi_2013} and as the latter was extracted experimentally in \cite{Rivera_2000} to be $\Omega_{exp} = 3000~1/s^2$, we can proof that the same $u_{rms}$ could only be reached due to an excessive energy input. The latter has to compensate the excessive viscous dissipation, which can be attributed to the excessive averaged enstrophy from the numerical simulation in the interval $I_T = [3;8.4]~s$, namely:
\begin{equation}
\Omega = \frac{1}{|I_T|} \int_{I_T} \frac{1}{A}\int_A \omega^2 d\mathbf{x}~dt \approx 8200~1/s^2~
\label{eq:Enstrophy}
\end{equation} 
Although the simulation was conducted with a spatial resolution of $\Delta l$ twice as fine as the experimentally extracted viscous friction scale $l_{\eta}$, we still obtain an excessive viscous dissipation by a factor of $\epsilon_{\nu}/\epsilon_{\nu,exp} \sim \Omega/\Omega_{exp} \approx 2.7$ with regard to the experiment. 
\\
In this sense, we have still performed a spatially underresolved direct numerical simulation of the turbulent fluid flow and the deficits of the pseudo-Lagrangian behaviour, as described in section \ref{sec:Motivation}, become obvious. Hence, it seems natural to evaluate the molecular stress tensor in Eq. (\ref{eq:MolecularStressTensor}) to locate the spatial regions where the pseudo-Lagrangian character becomes important.

\subsection{Molecular Stresses in Two-Dimensional Turbulence}

\begin{figure*}

\begin{subfigure}{.49\textwidth}
\centering
\includegraphics[clip, trim= 0.5cm 1.5cm 0.5cm 2.2cm, width=3.1in]{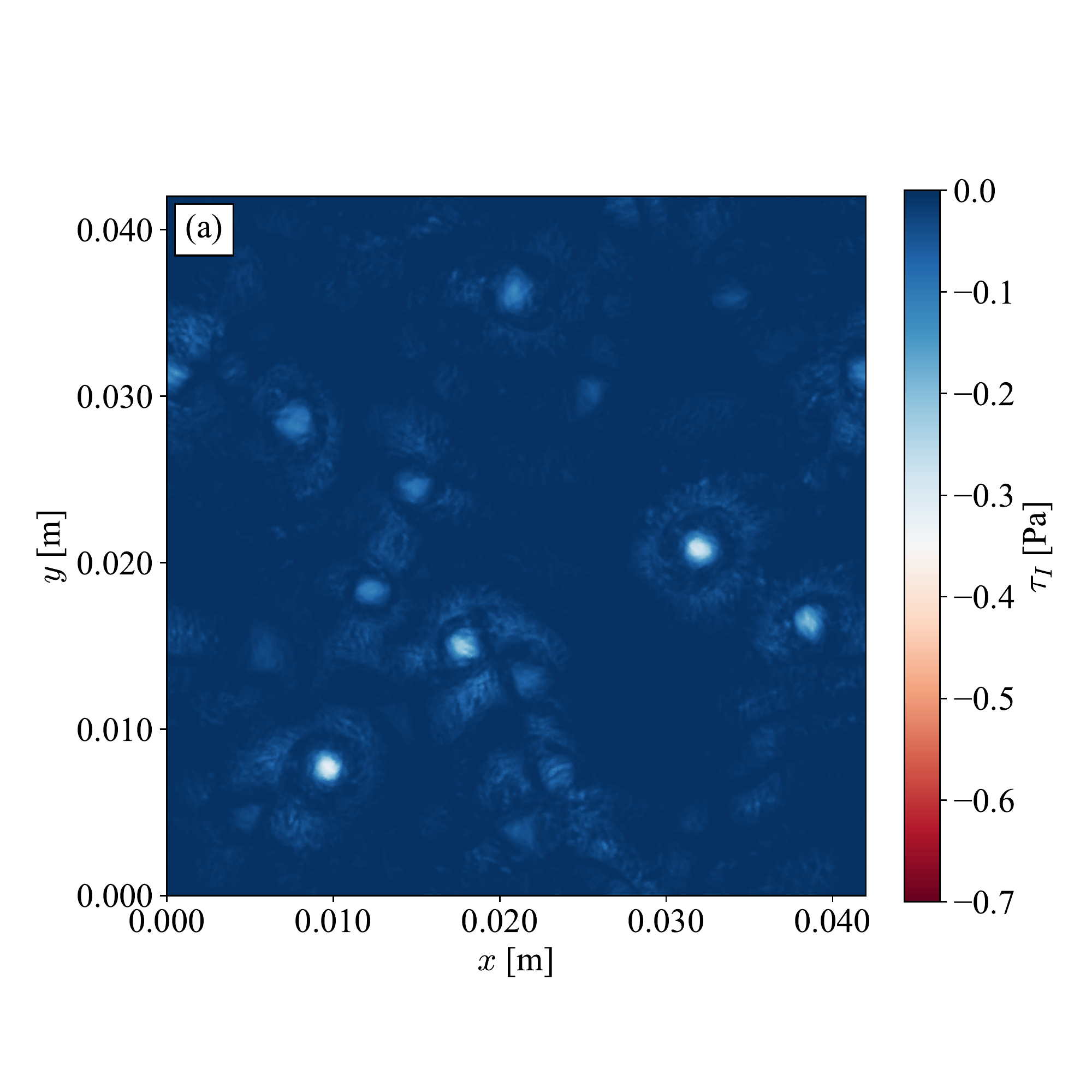}
%\caption{Snapshot of the first eigenvalue $\tau_{I}$.}
\label{fig:MolStressesA}
\end{subfigure}
\hfill
\begin{subfigure}{.49\textwidth}
\centering
\includegraphics[clip, trim= 0.5cm 1.5cm 0.5cm 2.2cm, width=3.1in]{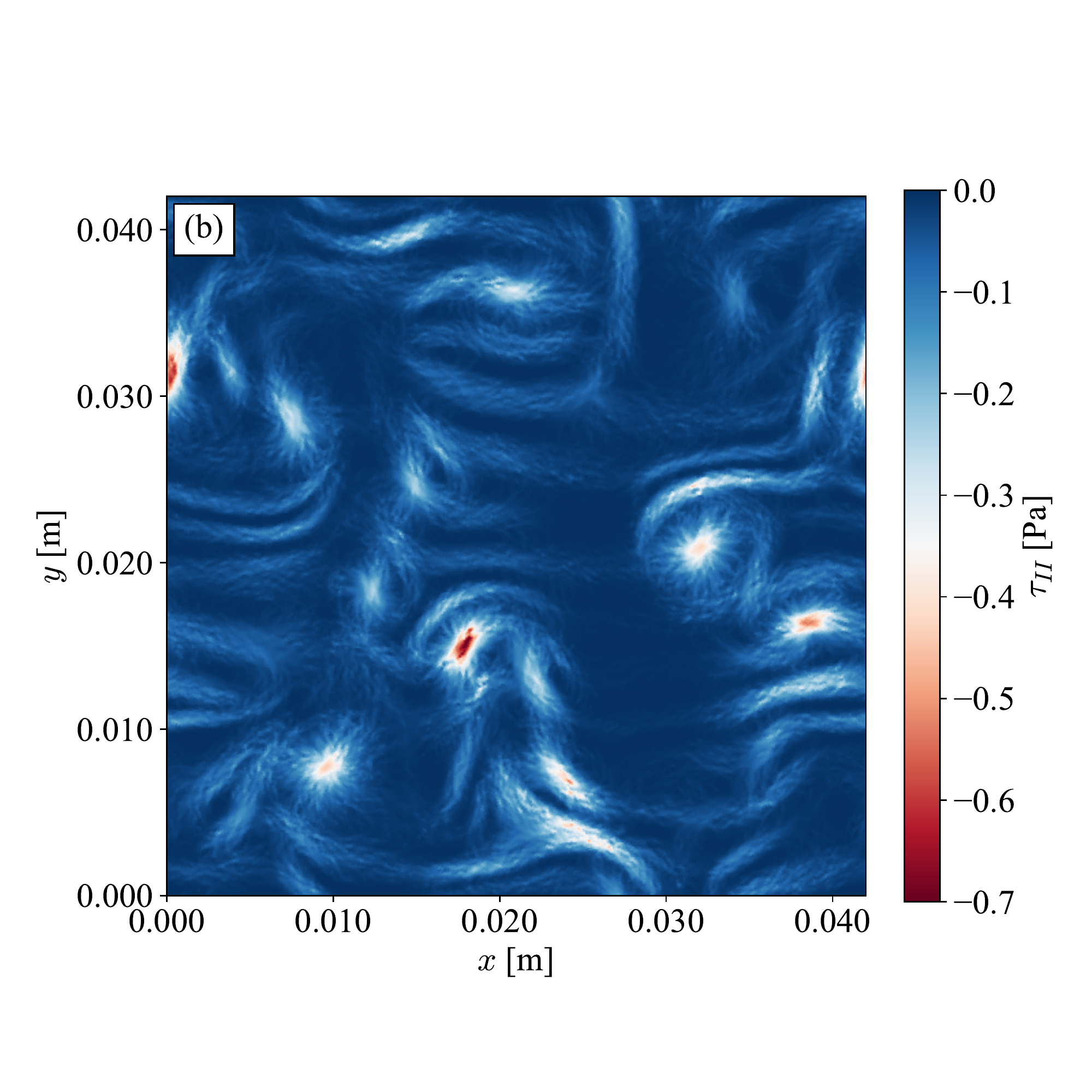}
%\caption{Snapshot of the second eigenvalue $\tau_{II}$.}
\label{fig:MolStressesB}
\end{subfigure}

\begin{subfigure}{.49\textwidth}
\centering
\includegraphics[clip, trim= 0.5cm 1.5cm 0.5cm 2.2cm, width=3.1in]{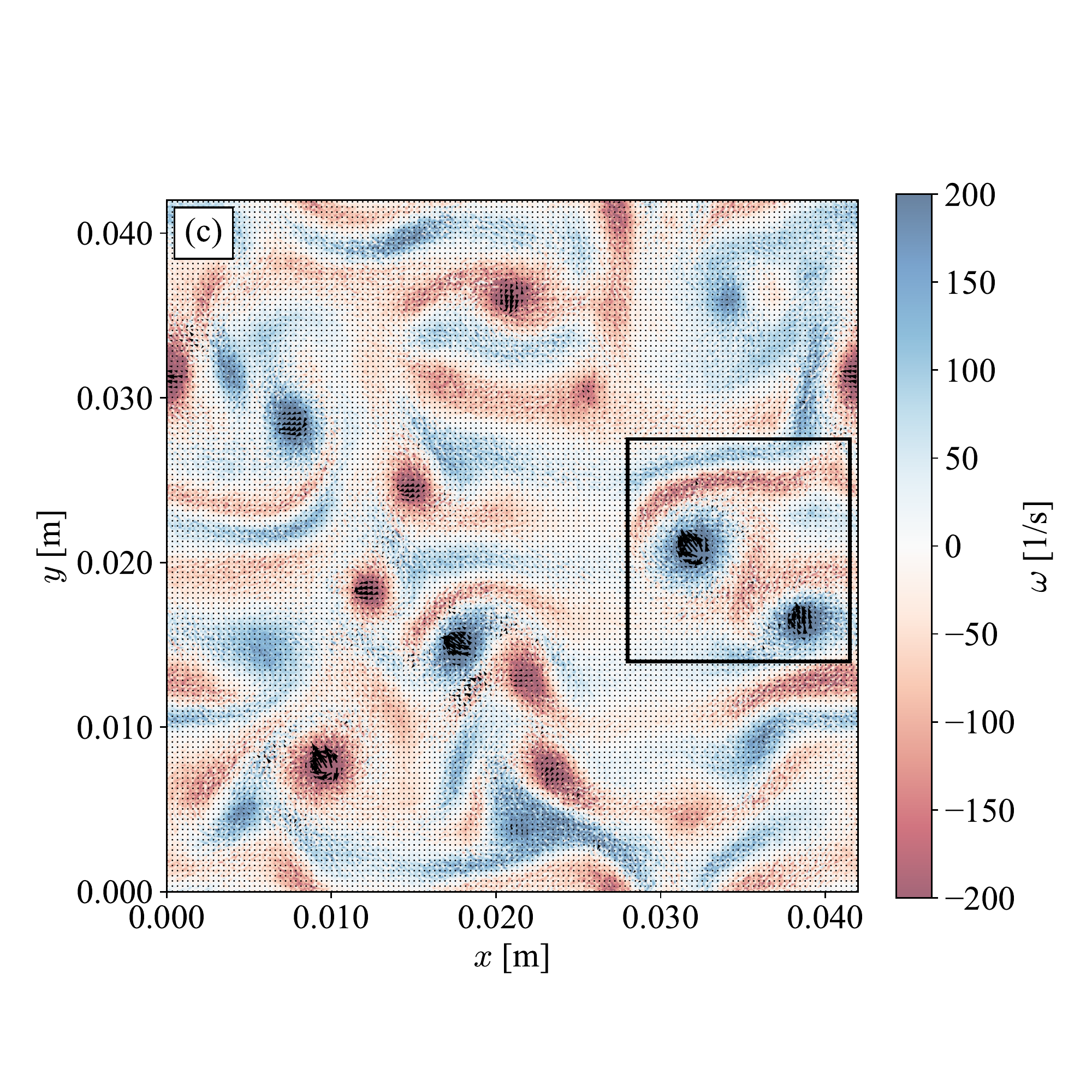}
%\caption{Snapshot of the first eigenvector $\mathbf{t}_I$ with region considered for detailed view. The corresponding vorticity field is superimposed in the background.}
\label{fig:MolStressesC}
\end{subfigure}
\hfill
\begin{subfigure}{.49\textwidth}
\centering
\includegraphics[clip, trim= 0.5cm 1.5cm 0.5cm 2.2cm, width=3.1in]{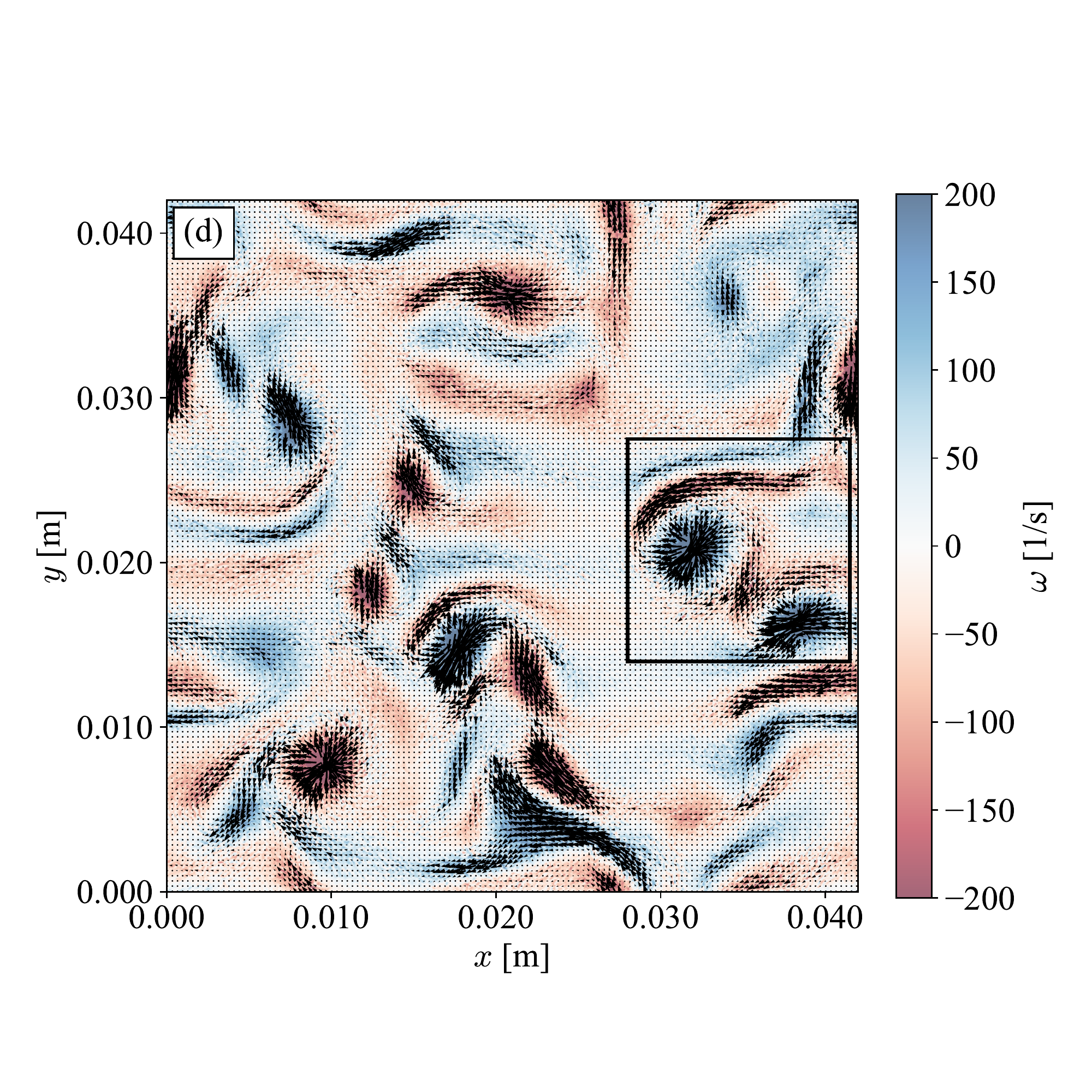}
%\caption{Snapshot of the second eigenvector $\mathbf{t}_{II}$ with region considered for detailed view. The corresponding vorticity field is superimposed in the background.}
\label{fig:MolStressesD}
\end{subfigure}

\begin{subfigure}{.49\textwidth}
\centering
\includegraphics[clip, trim= 0.5cm 1.5cm 0.5cm 2.2cm, width=3.1in]{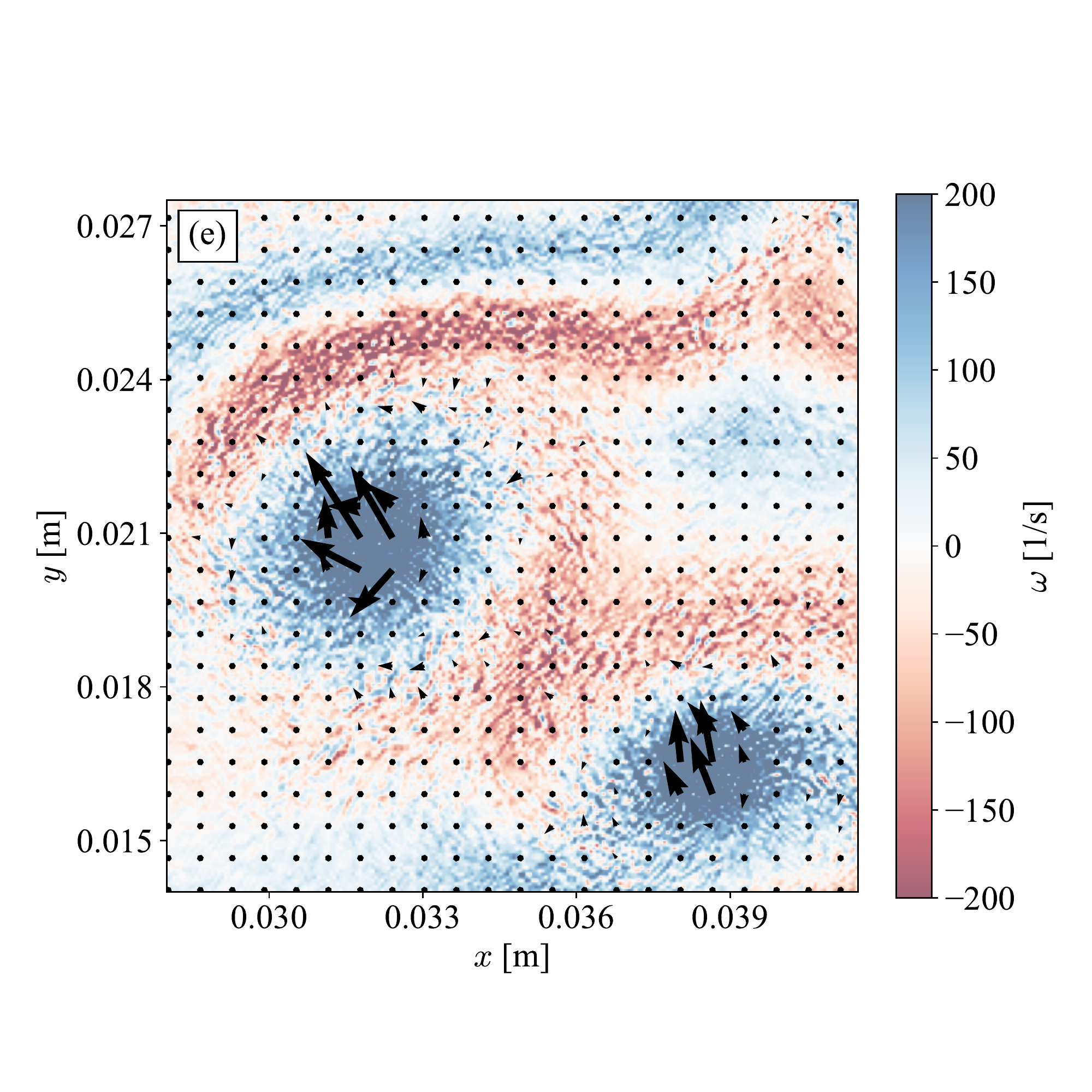}
%\caption{Detail of first eigenvector $\mathbf{t}_I$. The corresponding vorticity field is superimposed in the background.}
\label{fig:MolStressesE}
\end{subfigure}
\hfill
\begin{subfigure}{.49\textwidth}
\centering
\includegraphics[clip, trim= 0.5cm 1.5cm 0.5cm 2.2cm, width=3.1in]{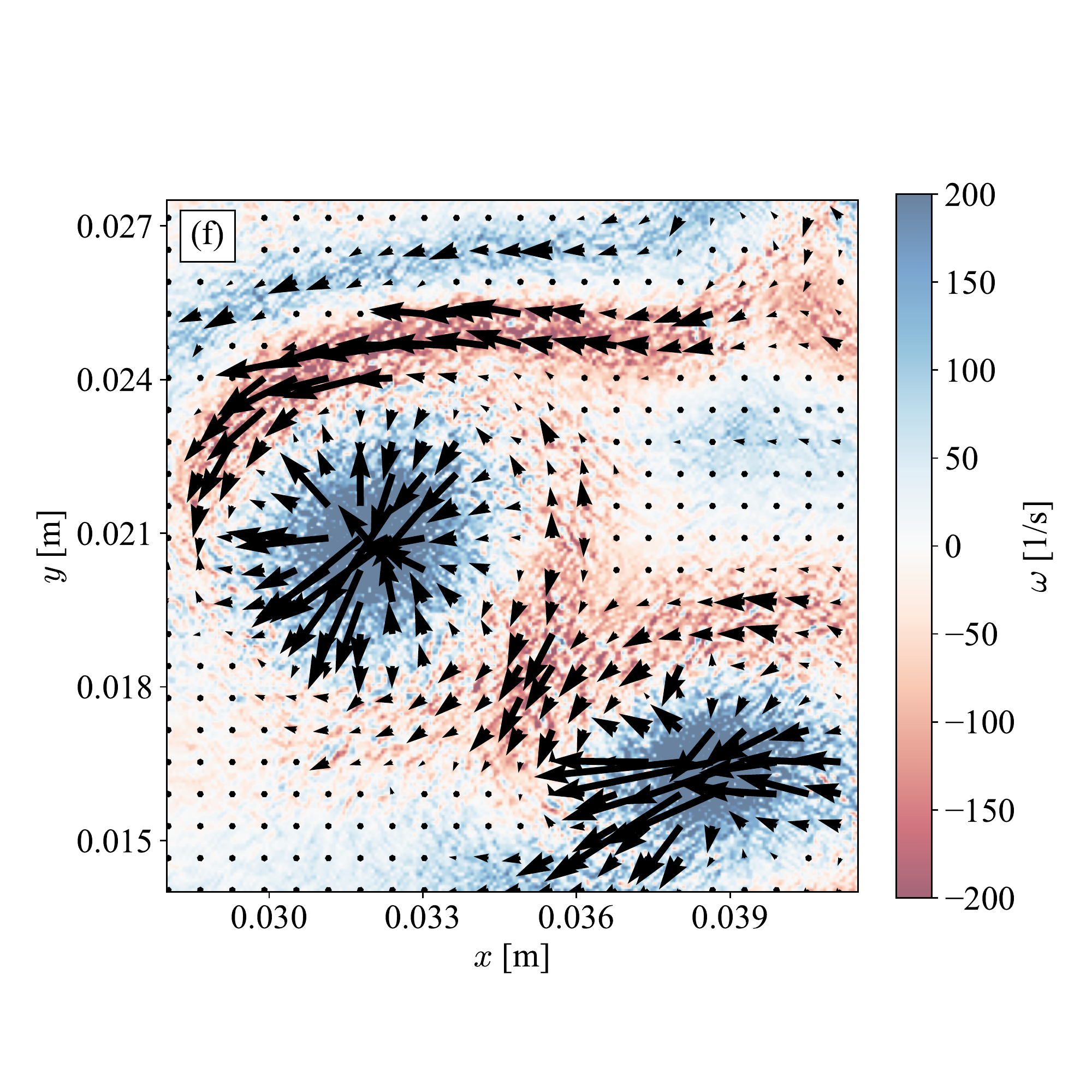}
%\caption{Detail of the second eigenvector $\mathbf{t}_{II}$. The corresponding vorticity field is superimposed in the background.}
\label{fig:MolStressesF}
\end{subfigure}
 
\caption{Visualization of the molecular stress tensor $\boldsymbol{\tau}_{mol}$ at time $t^*=4.1~s$ by eigendecomposition. For (d) - (f) the corresponding vorticity field is superimposed in the background. (a)  Snapshot of the first eigenvalue $\tau_{I}$. (b) Snapshot of the second eigenvalue $\tau_{II}$. (c) Snapshot of the first eigenvector $\mathbf{t}_I$ with region considered for detailed view. (d) Snapshot of the second eigenvector $\mathbf{t}_{II}$ with region considered for detailed view. (e) Detail of the first eigenvector $\mathbf{t}_I$. (f) Detail of the second eigenvector $\mathbf{t}_{II}$.}
\label{fig:MolStresses}
\end{figure*}

As the molecular stress described by $\boldsymbol{\tau}_{mol}$ in Eq. (\ref{eq:MolecularStressTensor}) varies in space and time, we present $\boldsymbol{\tau}_{mol}$ only for the turbulent snapshot depicted at time $t^* = 4.1~s$ in Fig. (\ref{fig:SnapShots}). The essence of the observations is not affected by this restriction. For the evaluation of $\boldsymbol{\tau}_{mol}$ in Eq. (\ref{eq:MolecularStressTensor}) each particle was first associated with a kernel element $K$ whose size is determined by the selected effective number of neighbors $N_{NGB} = 32$ \cite{Hopkins_2015}. Then, the corresponding kernel velocities $\mathbf{U}$ according to Eq. (\ref{eq:KernelVelocity}) were calculated to obtain the peculiar $\mathbf{w}_j$ velocities within $K$ as given by Eq. (\ref{eq:VDecomp}). These were finally utilized for the computation of the molecular stress $\boldsymbol{\tau}_{mol}$. Additionally, to shorten the description and reduce the amount of presented data, an eigendecomposition of $\boldsymbol{\tau}_{mol}$ was performed. In $\mathbb{R}^2$ this reads
\begin{equation}
\boldsymbol{\tau}_{mol}\mathbf{t}_{I,II} = \tau_{I,II} \mathbf{t}_{I,II}~,
\label{eq:Eigendecomposition}
\end{equation}
where $\tau_{I,II}$ denotes the first or second eigenvalue and $\mathbf{t}_{I,II}$ the first or second eigenvector. According to continuum mechanics, we will call the eigenvalues principal stresses and the eigenvectors principal directions \cite{Coman_2020}. The principal stresses in $\mathbb{R}^2$ are explicitly given by \cite{Lubliner_2019}:
\begin{equation}
\tau_{I,II} = \frac{tr(\boldsymbol{\tau}_{mol})}{2} \pm \sqrt{\left( \frac{\tau_{mol,11} - \tau_{mol,22}}{2} \right)^2 + \tau_{mol,12}^2}~.
\label{eq:Eigenvalues}
\end{equation}
In Eq. (\ref{eq:Eigenvalues}) the expression $tr(\boldsymbol{\tau}_{mol}) = \tau_{mol,11} + \tau_{mol,22}$ denotes the trace of the tensor field. By inserting Eq. (\ref{eq:Eigenvalues}) into Eq. (\ref{eq:Eigendecomposition}), the resulting linear system can be solved and the corresponding principal directions are derived as:
\begin{equation}
\mathbf{t}_{I,II} = \left( 1, ~ -\frac{\tau_{mol,11} - \tau_{I,II}}{\tau_{mol,12}}\right)^T~.
\label{eq:Eigenvectors}
\end{equation}
It should be noted that $\mathbf{t}_{I}$ and $\mathbf{t}_{II}$ are orthogonal to each other everywhere, which manifests in $\mathbf{t}_{I} \cdot \mathbf{t}_{II}=0$. We want to emphasize, that in the subsequent visualization this property is hardly perceptible. For visualization purposes, we further normalize the eigenvector fields with their Euclidean norm $||\mathbf{t}_{I,II}||_2$ and scale them with the conjugated eigenvalues in Eq. (\ref{eq:Eigenvalues}). This transformation reads:
\begin{equation}
\mathbf{t}_{I,II} \rightarrow \tau_{I,II}\frac{\mathbf{t}_{I,II}}{||\mathbf{t}_{I,II}||_2}~.
\label{eq:Scaling}
\end{equation}
The results of the eigendecomposition of $\boldsymbol{\tau}_{mol}$ at time $t^*$ are depicted in Fig. (\ref{fig:MolStresses}) and we will start with a discussion of the principal stresses.
\begin{figure*}[th]

\begin{subfigure}{.49\textwidth}
\centering
\includegraphics[clip, trim= 0.5cm 1.5cm 0.5cm 2.2cm, width=3.1in]{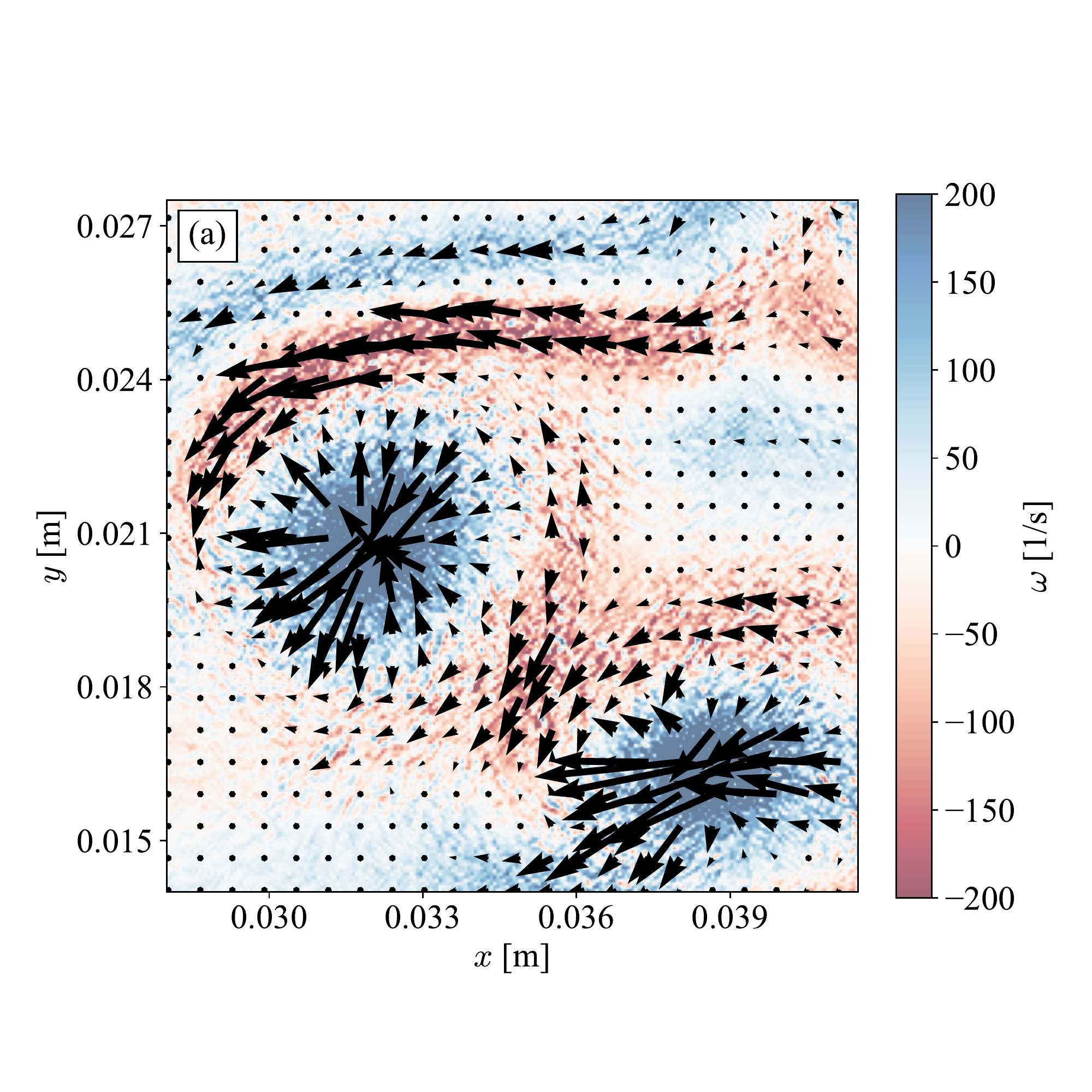}
%\caption{Detail of the second eigenvector $\mathbf{t}_{II}$.}
\label{fig:JuxtaA}

\end{subfigure}
\begin{subfigure}{.49\textwidth}
\centering
\includegraphics[clip, trim= 0.5cm 1.5cm 0.5cm 2.2cm, width=3.1in]{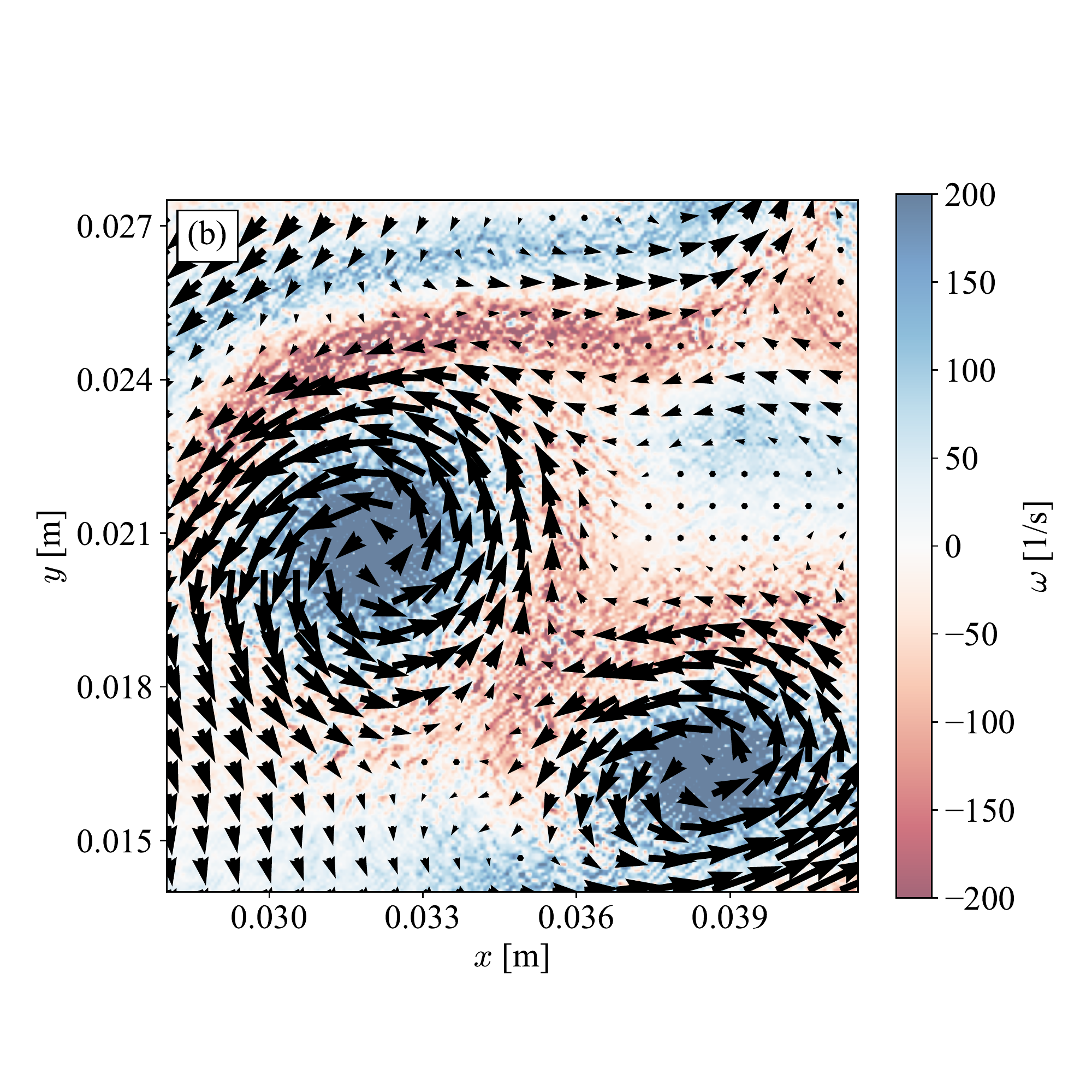}
%\caption{Detail of the velocity field $\mathbf{v}$.}
\label{fig:JuxtaB}
\end{subfigure}

\caption{Juxtaposition of (a) the detail of the second eigenvector $\mathbf{t}_{II}$ and (b) the detail of the velocity field $\mathbf{v}$ for the statistically steady turbulent flow field at time $t^*=4.1~s$. The corresponding vorticity field is superimposed in the background. A comparison reveals that the eigenvectors inside the vortex cores are dominantly oriented perpendicular to the flow field whereas on high vorticity filaments, in the vortex surrounding, the eigenvectors are dominantly oriented tangential in respect to the main flow.}
\label{fig:PreferentialFlow}
\end{figure*}\\
As can be seen in Fig. (\ref{fig:MolStresses}a), the first principal stress field $\tau_{I}$ is close to zero almost everywhere except where vortex cores are present. A comparison with the corresponding vorticity field in Fig. (\ref{fig:MolStresses}c) proves this statement. It gives an indication that the pseudo-Lagrangian character becomes sensible in the vicinity of vortices. This observation also holds true for the principal stress field $\tau_{II}$ in Fig. (\ref{fig:MolStresses}b). Again, the maximal principal stress magnitude coincides with the location of vortex cores, although it has to be highlighted that the magnitudes are significantly higher. On the contrary, the quantity $\tau_{II}$ has a significant, non-zero intensity in the regions of high vorticity filaments as well. Thus, we can conclude that molecular stresses $\boldsymbol{\tau}_{mol}$ are introduced everywhere in the flow domain where strong velocity gradients prevail. This is reasonable as the peculiar velocities $\mathbf{w}_j$ associated with a kernel element $K$ generally grow with increasing velocity gradients by definition of Eq. (\ref{eq:KernelVelocity}) and Eq. (\ref{eq:VDecomp}). Since $\mathbf{w}_j$ quadratically feeds back into $\boldsymbol{\tau}_{mol}$ in Eq. (\ref{eq:MolecularStressTensor}), the molecular stress grows significantly in regions of strong shear. Furthermore, the molecular stress in shear regions will be additionally increased by the agitation instability described by Basa et al. \cite{Basa_2009} in the WCSPH context. According to this instability, particles exposed to shear flow are not solely convected into the main flow direction but also exhibit a perpendicular movement, which causes irregular particle distributions. As a result, by implication of Eq. (\ref{eq:VolEst}), volume partition errors are generated, as well as thermal noise, which is responsible for further relative particle movement inside $K$. Additional non-zero $\mathbf{w}_j$ are triggered which consequently contribute to the molecular stress $\boldsymbol{\tau}_{mol}$.\\
Even more interesting insight into the physical nature of the molecular stress $\boldsymbol{\tau}_{mol}$ can be gained by the analysis of the principal directions as defined by Eq. (\ref{eq:Scaling}). So far, we could demonstrate, by means of the principal stress fields $\tau_{I}$ and $\tau_{II}$, that significant molecular stress is inherently linked to areas of high velocity gradients. Intriguingly, the orientation of the generated molecular stress is characteristic as well. In Fig. (\ref{fig:MolStresses}c) and Fig. (\ref{fig:MolStresses}d) the eigenvector fields $\mathbf{t}_{I}$ and $\mathbf{t}_{II}$, corresponding to Fig. (\ref{fig:MolStresses}a) and Fig. (\ref{fig:MolStresses}b), are visualized as black arrows. Both are superimposed on the vorticity field $\omega$. As the magnitude of the principal directions is proportional to $\tau_{I,II}$ according to Eq. (\ref{eq:Scaling}), it is comprehensible that in Fig. (\ref{fig:MolStresses}c) the eigenvector field $\mathbf{t}_{I}$ is hardy visible. Only a few vectors are apparent inside the vortex cores. On the other hand the eigenvector field $\mathbf{t}_{II}$ in Fig. (\ref{fig:MolStresses}d) is clearly visible in areas where the black vectors are concentrated. To reveal the characteristic structure of the eigenvector fields, we will subsequently focus on the black outlined region as highlighted in Fig. (\ref{fig:MolStresses}c) and Fig. (\ref{fig:MolStresses}d).  The outcome is visualized in Fig (\ref{fig:MolStresses}e) and Fig. (\ref{fig:MolStresses}f). As already stated, in Fig. (\ref{fig:MolStresses}e) only a few vectors are present in the vicinity of the vortex core. If we consider natural coordinates in each point of the field, two types of vectors can be distinguished. The first type is mainly oriented perpendicular to the main flow direction whereas the second type is mainly tangential to the main flow direction. In Fig. (\ref{fig:MolStresses}f) this observation is supported in a more obvious way as the magnitude of the eigenvector field $\mathbf{t}_{II}$ is much bigger due to $\tau_{II}$. Furthermore, it can be demonstrated by Fig. (\ref{fig:MolStresses}f) that not only do the vectors fields possess preferential directions but that the preferential directions are also linked to certain flow features. Vectors associated with high vorticity filaments are preferentially oriented tangential to the main flow direction in contrast to vectors associated with vortex spots which are mainly oriented perpendicularly to the main flow. This fact is highlighted in Fig. (\ref{fig:PreferentialFlow}) in which the detail of the eigenvector field $\mathbf{t}_{II}$ is juxtaposed in opposition to the vector field of the velocity $\mathbf{v}$. \\
We conclude that the implicit molecular stress in regions of strong shear is characterized by what we call shear modes and stretch modes. The former are aligned with the main flow direction whereas the latter are oriented perpendicular to it. Both are a consequence of the fact that for a given velocity gradient the local size of the kernel-based fluid element $K$ is too large, such that particles interact with each other in a detrimental way. \\
Since we could demonstrate in section \ref{sec:Convergence} that the molecular stress introduces an additional diffusional momentum transfer (Eq. (\ref{eq:KernelNavierStokes})), we suppose that the difference between these two modes is how the kinetic energy of the flow is redistributed. Depending on the local bending of the streamlines, two cases should be discriminated which are depicted in Fig. (\ref{fig:Modes}):
\begin{itemize}
\item Shear modes, associated with regions of moderate streamline bending, will redistribute the kinetic energy between parallel velocity components pointing into the main flow direction in order to homogenize the velocity field . 

\item Stretch modes, associated with regions of strong streamline bending, will redistribute the kinetic energy between orthogonal velocity components aligned with the principal directions given by Eq. (\ref{eq:Scaling}).
\end{itemize}

Hence, stretch modes could be a physical explanation for the fact that weakly-compressible particle based discretization methods are prone to excessive dissipation when decaying vortices are considered \cite{Hopkins_2015, Rossi_2015}.

\begin{figure*}[th]
\vspace{-0.4cm}
\begin{subfigure}{.49\textwidth}
\centering
\includegraphics[trim= 0cm 0cm 0cm 0cm, width=1.5in]{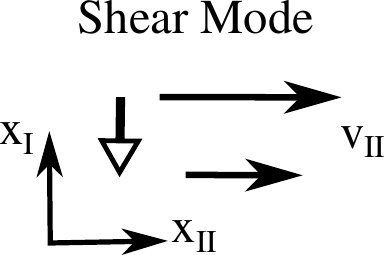}
%\caption{Illustration of the shear mode. Two parallel velocity vectors in the eigenbasis defined by Eq. (\ref{eq:Scaling}) are depicted. The molecular energy transfer is indicated by a white arrow.}
\label{fig:ModeA}

\end{subfigure}
\begin{subfigure}{.49\textwidth}
\centering
\includegraphics[trim= 0cm 0cm 0cm 0cm, width=1.5in]{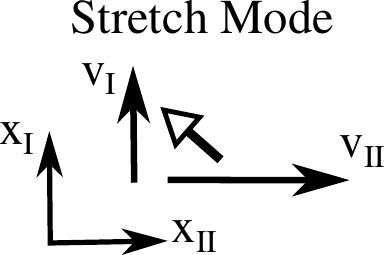}
%\caption{Illustration of the stretch mode. Two orthogonal velocity vectors in the eigenbasis defined by Eq. (\ref{eq:Scaling}) are depicted. The molecular energy transfer is indicated by a white arrow.}
\label{fig:ModeB}
\end{subfigure}

\caption{Schematic illustration of the modes extracted from the molecular stress tensor $\boldsymbol{\tau}_{mol}$. Shear mode: Two parallel velocity vectors in the eigenbasis defined by Eq. (\ref{eq:Scaling}) are depicted. The molecular energy transfer is indicated by a white arrow. Stretch mode: Two orthogonal velocity vectors in the eigenbasis defined by Eq. (\ref{eq:Scaling}) are depicted. The molecular energy transfer is indicated by a white arrow.}
\label{fig:Modes}
\end{figure*}

\subsection{Molecular Stresses as Basis for Conceptual Improvements}
\label{subsec:Paths}

Although details of the energy transfer with regard to molecular stresses are not fully understood, it is evident at this point that $\boldsymbol{\tau}_{mol}$ proved to be an appropriate indicator to pin down the pseudo-Lagrangian character. Hence, we strive for an answer on how $\boldsymbol{\tau}_{mol}$ can be utilized to mitigate the problems associated with the pseudo-Lagrangian character. At this stage, we see two promising research paths for conceptual improvements:
\\
\\
\emph{Subkernel-scale model:} Assuming that the utilized local kernel elements are appropriate representatives of local fluid elements, one possible way to eliminate the pseudo-Lagrangian character could be to explicitly consider $div(\boldsymbol{\tau}_{mol})$ in the methods of interest. Then, the term would serve as a subkernel-scale model in accordance to subgrid-scale models deployed in turbulence modelling. Recalling that $\mathbf{a} \approx \sum_{j=1}^{N_{NGB}} \mathbf{a}_jW_h(\mathbf{x}-\mathbf{x}_j) V_j$ is denoting the numerically approximated acceleration at position $\mathbf{x}$, we rearrange Eq. (\ref{eq:KernelNavierStokes}) as follows:
\begin{equation}
\rho \mathbf{a} = - div( \boldsymbol{\tau}_{mol}) + \rho \frac{d\mathbf{U}}{dt}~.
\label{eq:Subkernel_Model}
\end{equation}
From Eq. (\ref{eq:Subkernel_Model}) it can be concluded that it may be interesting to investigate how particle methods perform if the kernel averaged velocity $\mathbf{U}$ is algorithmically utilized and $- div( \boldsymbol{\tau}_{mol})$ considered as an extra stress term. \\
If we subsequently restrict ourselves to the extra stress term in Eq. (\ref{eq:Subkernel_Model}) and decompose the molecular stress tensor $\boldsymbol{\tau}_{mol}$ into its spherical and deviatoric part, namely $sp\{\boldsymbol{\tau}_{mol}\}$ and $dev\{\boldsymbol{\tau}_{mol}\}$, one finds in $\mathbb{R}^2$:
\begin{eqnarray}
- div( \boldsymbol{\tau}_{mol}) &=& - div \left( sp\{\boldsymbol{\tau}_{mol}\} + dev\{\boldsymbol{\tau}_{mol}\} \right) \nonumber \\
 &=& - div \left( \frac{tr(\boldsymbol{\tau}_{mol})}{2} + dev\{\boldsymbol{\tau}_{mol}\} \right)~. 
\label{eq:TauMolDecomp}
\end{eqnarray}
Surprisingly, if only the spherical part of $\boldsymbol{\tau}_{mol}$ is considered in Eq. (\ref{eq:TauMolDecomp}), then the extra stress term reads:
\begin{equation}
- div \left( \frac{tr(\boldsymbol{\tau}_{mol})}{2}\right) = \frac{1}{2} \sum_{j=1}^{N_{NGB}} M \nabla(\mathbf{w}_j^2 W_h(\mathbf{x}-\mathbf{x}_j))~.
\label{eq:LANSEquivalent}
\end{equation} 
The result in Eq. (\ref{eq:LANSEquivalent}) is very similar to the extra stress term developed by Monoghan \cite{Monoghan_2017} within the SPH-$\epsilon$ turbulence model, although obtained in a completely different way. While the model of Monoghan emerges from the Lagrangian of the particle system \cite{Monoghan_2017} by means of a top-down approach, our model is a consequence of a bottom-up approach by application of the Hardy theory from NEMD \cite{Hardy_1982}. Compared to the extra stress term in \cite{Monoghan_2017}, the term in Eq. (\ref{eq:LANSEquivalent}) does not contain a numerical parameter $\epsilon \in [0,1]$ and gradients of the quadratic relative velocity $\mathbf{w}_j^2$ are considered as well. It is hard to believe that this similarity is just a coincidence. Furthermore, our proposed model would contain information about the deviatoric part of $\boldsymbol{\tau}_{mol}$ as well (Eq. (\ref{eq:TauMolDecomp})) and was not explicitly developed for turbulence modelling.
\\
\\
\emph{Dynamic Anisotropic Kernel Adjustment:} Another promising research path emerging from the utilization of the molecular stress $\boldsymbol{\tau}_{mol}$ could be to dynamically adjust the shape of the kernel. Instead of using an isotropic spherical kernel in a strongly anisotropic flow, which causes pseudo-Lagrangian behaviour, it could be attractive to utilize the scaled principle directions (Eq. (\ref{eq:Scaling})) to adapt the kernel in a Lagrangian manner. An initially spherical kernel-based fluid element would then be deformed into an ellipsoid according to the flow as sketched in Fig. (\ref{fig:GenericFlow}). Hence, the pseudo-Lagrangian character would be mitigated in its roots in contrast to the proposed subkernel-scale model strategy. The idea is not new in general and was already introduced by Shapiro et al. \cite{Shapiro_1996} with the only difference that an eigendecomposition of the velocity field Jacobian $\mathbf{v}'$ rather than $\boldsymbol{\tau}_{mol}$ was proposed for the kernel adjustment. We see a potential benefit in the fact that for the calculation of $\boldsymbol{\tau}_{mol}$ no estimates of the spatial derivatives are required which, at least in SPH, can be heavily affected by noise. Although introducing computational overhead for the kernel adjustment, there is the chance that this modification will improve the performance. As explained by Shapiro et al. \cite{Shapiro_1996}, the benefit of a Lagrangian adjustment of the kernel is that a costly nearest neighbor particle search does not have to be conducted every time step. It would be interesting to see how a specific implementation of such an anisotropic kernel adjustment performs compared to the conventional reference.

\section{Conclusion}

In this work, isothermal kernel transport equations were derived that describe the dynamics of kernel-based fluid elements in weakly-compressible particle based discretization methods. Therefore, we applied the theory of Hardy \cite{Hardy_1982} from NEMD, motivated by the fact that these methods suffer from pseudo-Lagrangian behaviour. Interestingly, the outcome of this bottom-up approach was that the resulting kernel transport equations contain an additional molecular stress term $\boldsymbol{\tau}_{mol}$. We could explain that these stresses are associated with the thermal noise triggered by volume partition errors as well as the isotropic, spherical kernels utilized. Furthermore, it was demonstrated by means of a generic flow example that the stresses generate forces that act opposite to the actual Lagrangian deformation of the kernel-based fluid element $K$. Mathematically, these molecular forces are driven by the curvature of the velocity field and, hence, introduce a diffusional momentum transfer in the kernel transport equations. \\
Another major result, which could be valuable for theorists, is that convergence of weakly-compressible particle based discretization schemes can not only be physically interpreted as the elimination of the kernel low-pass filtering but also the elimination of implicit molecular stresses. \\
Finally, we have analyzed the ability of the molecular stress tensor $\boldsymbol{\tau}_{mol}$ as a quality indicator for a numerical dataset acquired with the MFM method \cite{Hopkins_2015}. Therefore, we numerically reproduced the two-dimensional turbulent flow of Rivera and Wu \cite{Rivera_2001} with the open-source code GIZMO \cite{Hopkins_2017}. Although the same level of averaged turbulent kinetic energy could be reached, this was only possible due to excessive energy input which compensates the excessive viscous dissipation. We believe that this excessive viscous dissipation is implictly related to the molecular stress. An eigenvalue decomposition of $\boldsymbol{\tau}_{mol}$ revealed that the numerical approximation in regions of high shear and vortical structures is prone to errors. This is well-known in the community \cite{Vogelsberger_2012, Zhu_2015, Hopkins_2015, Bauer_2012, Rossi_2015, Ellero_2010, Basa_2009}. Additionally, we were able to identify that $\boldsymbol{\tau}_{mol}$ is characterized by shear and stretch modes mainly aligned or perpendicular to the main flow. At the current stage, we presume that the latter can be distinguished by the way in which the kinetic energy is redistributed.

\appendix

\section{Property of Peculiar Momenta}
\label{sec:Appendix1}

In Eq. (\ref{eq:PeculiarMomenta}) it was stated that the kernel associated sum of the peculiar momenta vanishes identically. Here, a short reasoning is given. \\
Combining Eq. (\ref{eq:VDecomp}) and the lhs of Eq. (\ref{eq:PeculiarMomenta}), as well as taking into account that $M$ and $\mathbf{U}$ are independent of the summation index $j$, we can write:
%\begin{widetext}
\begin{eqnarray}
&&\sum_{j=1}^{N_{NGB}} M \mathbf{w}_j W_h(\mathbf{x}-\mathbf{x}_j) = \sum_{j=1}^{N_{NGB}} M (\mathbf{v}_j - \mathbf{U}) W_h(\mathbf{x}-\mathbf{x}_j) \nonumber \\
&&= M\sum_{j=1}^{N_{NGB}}  \mathbf{v}_j W_h(\mathbf{x}-\mathbf{x}_j) - M \mathbf{U} \sum_{j=1}^{N_{NGB}} W_h(\mathbf{x}-\mathbf{x}_j) \nonumber \\
&&= M \sum_{j=1}^{N_{NGB}} W_h(\mathbf{x}-\mathbf{x}_j) \left( \frac{\sum_{j=1}^{N_{NGB}} \mathbf{v}_j W_h(\mathbf{x}-\mathbf{x}_j)}{\sum_{j=1}^{N_{NGB}} W_h(\mathbf{x}-\mathbf{x}_j)} - \mathbf{U} \right) \nonumber \\
&& = \mathbf{0} \quad \qedsymbol \qquad
\label{eq:Proof}
\end{eqnarray}
%\end{widetext}
The second set of parentheses in the last part of Eq. (\ref{eq:Proof}) vanishes identically as its first term is exactly equal to the definition of the kernel velocity in Eq. (\ref{eq:KernelVelocity}). Thus, the statement in Eq. (\ref{eq:PeculiarMomenta}) is true.

\section{Gradient Approximation of Peculiar Velocities}
\label{sec:Appendix2}

Here we want to demonstrate that the peculiar velocity components $w_{j,\alpha}$ in Eq. (\ref{eq:VDecomp}), associated to a kernel element $K$ centered at $\mathbf{x} \in \mathbb{R}^3$, can be approximately represented by $\nabla v_\alpha(\mathbf{x})$. The index $\alpha$ denotes a spatial index. \\
We start with the utilization of the weakly-compressible character in Eq. (\ref{eq:KernelVelocity}). Then $V \approx V_j$ and we can write for the component $\alpha$ of the kernel velocity:
\begin{eqnarray}
U_\alpha := \frac{\sum_{j=1}^{N_{NGB}} v_{j,\alpha} W_h(\mathbf{x}-\mathbf{x}_j)}{\sum_{j=1}^{N_{NGB}} W_h(\mathbf{x}-\mathbf{x}_j)} \nonumber \\ 
\approx \sum_{j=1}^{N_{NGB}} v_{j,\alpha} W_h(\mathbf{x}-\mathbf{x}_j) V_j~.
\label{eq:AppendixWeaklyCompressible}
\end{eqnarray}
Additionally, we linearize the particle velocity $v_{j,\alpha}$ around the center $\mathbf{x}$. If we abbreviate $\mathbf{x}_j-\mathbf{x} =: \Delta \mathbf{x}_j$, we find:
\begin{equation}
v_{j,\alpha} \approx v_\alpha(\mathbf{x}) + \nabla v_\alpha(\mathbf{x}) \cdot \Delta \mathbf{x}_j~.
\label{eq:AppendixLinearization}
\end{equation}
Considering that for the kernel moments
\begin{eqnarray}
&\sum_{j=1}^{N_{NGB}} W_h (\mathbf{x}-\mathbf{x}_j)V_j \approx 1, \nonumber \\
&\sum_{j=1}^{N_{NGB}} \Delta \mathbf{x}_j W_h(\mathbf{x}-\mathbf{x}_j)V_j \approx \mathbf{0}~,
\label{eq:AppendixKernelMoments}
\end{eqnarray}
the peculiar velocity components in Eq. (\ref{eq:VDecomp}) can be approximated by combination of Eq. (\ref{eq:VDecomp}) with Eq. (\ref{eq:AppendixWeaklyCompressible}), Eq. (\ref{eq:AppendixLinearization}) and Eq. (\ref{eq:AppendixKernelMoments}). This results in the following gradient approximation:
\begin{equation}
w_{j,\alpha} \approx \nabla v_\alpha(\mathbf{x}) \cdot \Delta \mathbf{x}_j~.
\label{eq:AppendixGradientApproximation}
\end{equation}
As a consequence of Eq. (\ref{eq:AppendixGradientApproximation}), the peculiar velocities can only change in the spatial coordinate $\mathbf{x}$ if the particle velocity component $v_\alpha(\mathbf{x})$ has a non-zero curvature. For linear $v_\alpha(\mathbf{x})$ the gradient $\nabla v_\alpha(\mathbf{x})$ would be constant and thus $w_{j,\alpha} = const.$ for the same $\Delta \mathbf{x}_j$. Hence, without velocity noise on the particle level, the term $div(\boldsymbol{\tau}_{mol})$ in Eq. (\ref{eq:KernelNavierStokes}) would vanish.

\bibliographystyle{apsrev4-1} % Tell bibtex which bibliography style to use
\bibliography{xampl} % Tell bibtex which .bib file to use (this one is some example file in TexLive's file tree)

\end{document}